\documentclass[onecolumn, usenatbib]{mnras}
\usepackage{savesym}
\usepackage{graphicx}
\usepackage{longtable}
\usepackage{changepage}

\expandafter\let\csname equation*\endcsname\relax
  \expandafter\let\csname endequation*\endcsname\relax 
\usepackage{subfig}
\usepackage{amsmath}
\usepackage{amssymb}
\usepackage{verbatim}
\usepackage[yyyymmdd,hhmmss]{datetime}
\usepackage{array}
\usepackage{times}
\usepackage{xcolor}
\DeclareRobustCommand{\DE}[3]{#2}
\let\DEthebibliography\thebibliography
\def\thebibliography{\DeclareRobustCommand{\DE}[3]{##3}\DEthebibliography}

\title[Collisions with TDE disks]{Collisions with tidal disruption event disks: implications for quasi-periodic X-ray eruptions }
\author [Andrew Mummery]{Andrew Mummery$^1$\thanks{E-mail:
andrew.mummery@physics.ox.ac.uk} \\
$^1$Oxford Theoretical Physics, Beecroft Building,  Clarendon Laboratory, Parks Road, Oxford, OX1 3PU, United Kingdom 
}

\date{}

\begin{document}

\maketitle

\begin{abstract}
A popular class of models for interpreting quasi-periodic X-ray eruptions from galactic nuclei (QPEs) invoke collisions between an object on an extreme mass ratio inspiral (EMRI) and an accretion disk around a supermassive black hole. There are strong links between QPE systems and those disks which formed following a tidal disruption event (TDE), and at least two events (AT2019qiz and AT2022upj) are known to have occurred following an otherwise typical TDE. We show that the fact that these disks were formed following a TDE strongly constrains their properties, more so than previous models have assumed. Models based on steady-state AGN-like disks have mass contents which grow strongly with  size $M_{\rm disk}\propto R_{\rm out}^{7/2}$ and do not conserve the  mass or angular momentum  of the disrupted star. A very different scaling must be satisfied by a TDE disk in order to conserve the disrupted stars angular momentum, $M_{\rm disk} \propto R_{\rm out}^{-1/2}$. These constraints substantially change the predicted scaling relationships between QPE observables (luminosity, duration, energy, temperature) and the QPE period. They also allow QPE observables to be written in terms of the properties of the two stars assumed to be involved (the one tidally disrupted and the one on an EMRI), making plausibility tests of these models possible. We show that these modifications to the disk structure imply that (i) QPEs cannot be powered by collisions between an orbiting black hole and a TDE disk, (ii) QPEs also cannot be powered by collisions between the  surface of a stellar EMRI and a TDE disk. A framework in which the collisions are between a TDE disk and a star which has puffed up to fill its Hills sphere with a trailing debris stream (as seen in recent simulations) cannot be ruled out from the data, and should be the focus of further study. 
\end{abstract}

\noindent

\section{Introduction}
The discovery in 2019 of luminous ($\sim 10^{43}$ erg/s), quasi-periodic (repeating every $\sim 8$ hours) eruptions of hot ($kT_{\rm obs} \sim100$ eV) X-ray emission from the galactic nucleus of GSN 069 \citep{Miniutti2019} initiated the study of so-called quasi-periodic X-ray eruptions (QPEs). Since this first discovery,  9 more sources \citep{Giustini2020,Arcodia2021,Chakraborty2021, Quintin2023, Arcodia2024a,Guolo2024, Nicholl24, Chakraborty25, Hernandez25}. have been discovered, all broadly sharing the same observed characteristics. The sources now span recurrence periods of $\sim 2-72$ hours for ``classical'' QPEs, with a more atypical source having a recurrence time of 22 days \citep{Guolo2024}. 

Among many models, one class \citep[e.g.,][]{Dai2010,Xian2021,Sukova2021,Linial2023, Franchini2023, Tagawa2023, Vurm24, Yao24} invoking collisions between an object (either a star or an intermediate mass black hole) on an extreme mass ratio inspiralling (EMRI) orbit and an accretion flow around a supermassive black hole, are proving popular. It is the purpose of this Paper to examine these models in more detail following the discovery of long-period (i.e., $P_{\rm QPE} \sim {\cal O}({\rm days})$) QPEs at late times in AT2019qiz \cite{Nicholl19, Nicholl24} and AT2022upj \citep{Chakraborty25}, which before the detection of QPE flares were both otherwise ordinary tidal disruption events (TDE).  

Prior to AT2019qiz and AT2022upj there had been a number of previous observational hints that QPEs and TDEs were linked. They preferentially inhabit the same types of host galaxies \citep{Wevers2022}, and GSN 069 had been fading in a TDE-like manner for the 10 years prior to its discovery as a QPE source \citep{Miniutti2023}, as has a second QPE source \citep{Arcodia2024a}. Detailed spectral modelling has now confirmed that GSN 069 is a TDE \citep{Guolo25}. Two observed TDEs have exhibited X-ray flares consistent with individual eruptions \citep{Chakraborty2021,Quintin2023}. The discovery of QPEs in AT2019qiz and AT2022upj  confirms this link, and allows much more detailed probes of the physics of these orbiter-disk collision models to be performed. 

Within the framework of attempting to explain QPE properties as disk-EMRI collisions, it is furthermore rather natural to expect short period ($\sim 1-100$ hour) systems to be only present in new (i.e., TDE produced) accretion flows, as a radially extended and very massive AGN disk would likely destroy the star (through repeated collisions over a long period) before it could reach small enough radii to produce such short period flares. Modeling the specific properties of new, TDE-produced, disks is therefore very important to this paradigm. 

Each of the disk-orbiter model currently put forward in the literature either do not explicitly model the disk \citep[e.g.,][i.e., they simply assume that if a collision happens it will produce observable emission, independent of disk properties]{Dai2010,Xian2021, Franchini2023}, or invoke a standard steady-state AGN-like \cite{SS73} model for the quiescent disk \citep[e.g.,][]{Linial2023, Vurm24, Yao24}. While a natural starting point, these models are inappropriate for comparing to sources which follow a TDE, as AGN models have disk masses $M_{\rm disk}$ which grow strongly with radius $M_{\rm disk} \propto R_{\rm out}^{7/2}$, something which cannot be satisfied by a TDE disk as it would not conserve mass (which is bounded by the mass of the incoming star) or angular momentum (which is bounded by the angular momentum of the incoming stars orbit). Indeed, it will be shown in this paper that the maximum mass which is consistent with angular momentum conservation must fall as $M_{\rm disk, max} \propto R_{\rm out}^{-1/2}$ in a TDE disk.

\begin{figure}
    \centering
    \includegraphics[width=0.49\linewidth]{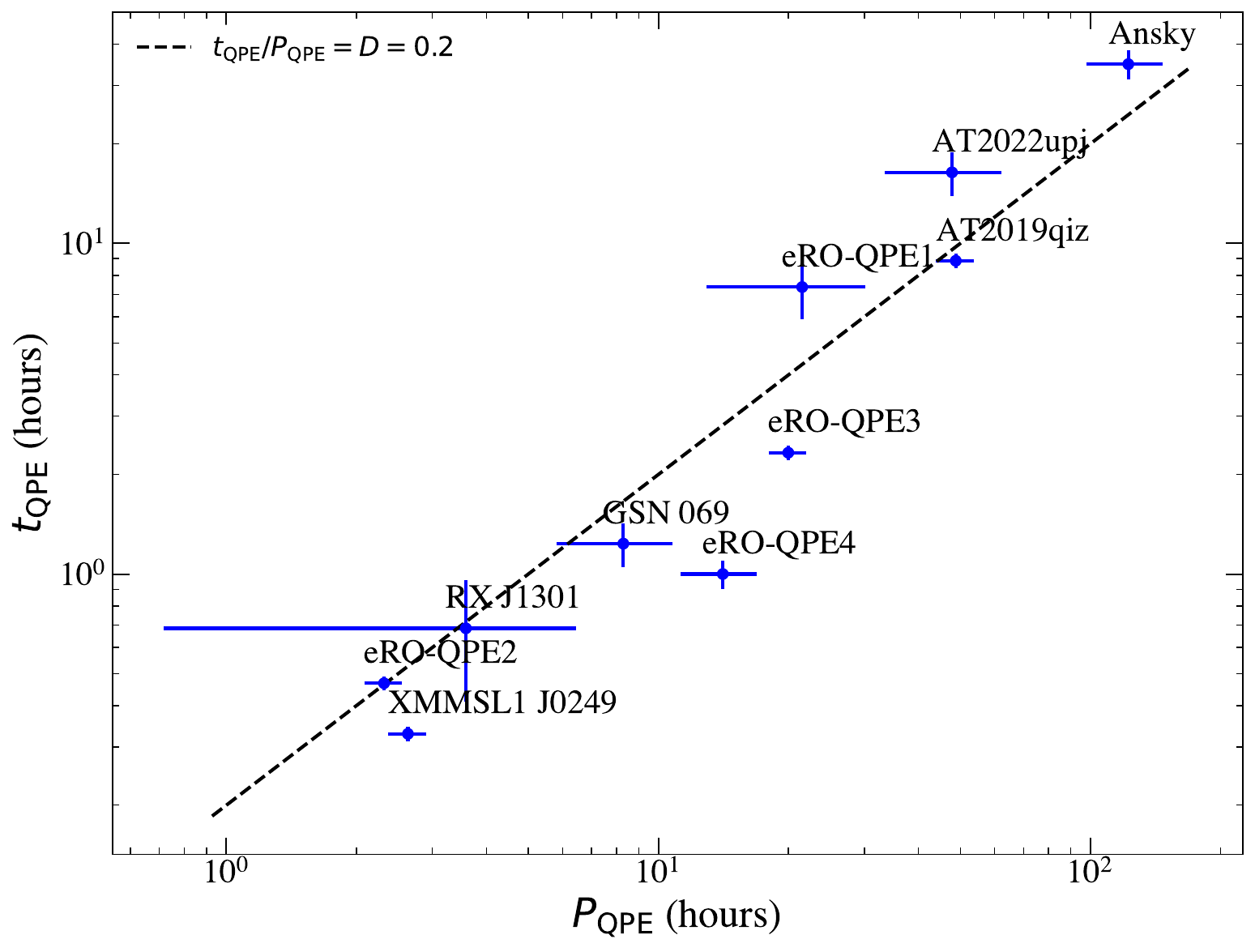}
    \includegraphics[width=0.49\linewidth]{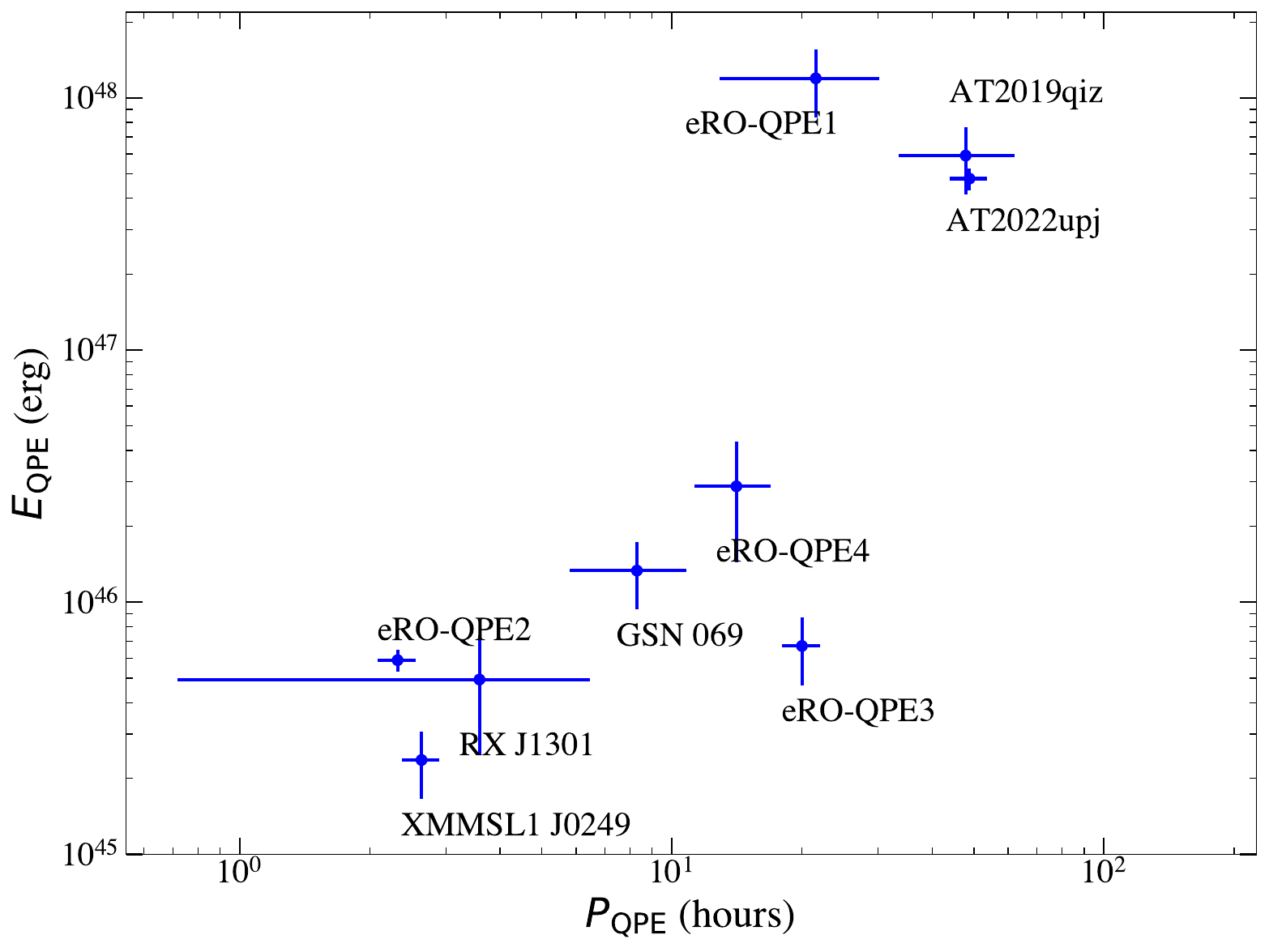}
    \caption{Two clear trends observed across the QPE population, whereby the length of the flares ($t_{\rm QPE}$) and the energy radiated in the flares (estimated here using $E_{\rm QPE} \approx L_{\rm bol, peak} \times t_{\rm QPE}$) correlate strongly with the time between flares ($P_{\rm QPE}$). The correlation  between duration and time between flares is especially tight, and satisfies a constant duty cycle relationship with $D \approx 0.2$ (black dashed curve). The energy in the QPE flares approximately satisfies $E_{\rm QPE} \propto P_{\rm QPE}^{5/3}$, although a simple fit has large associated errors for a small sample.  }
    \label{fig:data}
\end{figure}

We will show in this Paper that this quantitative change has serious implications for the TDE disk-EMRI collision model. Indeed, this bounded mass budget means that the longer one makes the time between collisions (i.e., the larger the size of the disk required for an interception with an orbiting body), the less mass one will collide with and therefore the less energy  that will be available to be radiated (the body doing the collisions will also be orbiting more slowly at this larger radii; all of these arguments will be made quantitative in the course of this paper). This is in  contrast with the observed correlation between time between collisions and radiated energy \citep[Figure \ref{fig:data}, with data taken form][]{Miniutti2019, Giustini2020,Arcodia2021,Chakraborty2021, Quintin2023, Arcodia2024a,Guolo2024, Nicholl24, Chakraborty25, Hernandez25}. The mass which the object collides with is also going to set the optical depth of the shocked gas, and therefore the timescale over which the photons escape the shocked debris (and therefore likely the timescale for the flare to evolve). Again, less mass being available to collisions at larger orbital radii will mean the photons should escape quicker from the gas cloud for larger QPE periods, also in contention with the data (Figure \ref{fig:data}).

We shall show in this paper that an immediate consequence of this dropping mass content in TDE disks which have spread to larger radii is that intermediate mass black holes (IMBHs) cannot possibly be the objects which collide with a disk to produce QPEs, as to sweep up enough of this low density TDE disk  into their Bondi radius  their masses  would have be so high that they themselves would accrete $\sim 0.1-10\%$ of the disk on each crossing (as well as likely imprinting other clear effects onto the original TDE emission). Similarly, it is not possible for the collision to be between the surface of a star and the disk, which simply cannot collide with enough material to produce the observed QPE energies. 

The only remaining region of parameter space left available to the QPE collision paradigm, we shall show, is for collisions between an extended structure formed of stellar debris and a TDE disk. To explain the most energetic QPEs this structure would have to have a width equal to $\sim$ the Hills radius of the EMRI in the disk plane, and a vertical extent which is even larger (i.e., a debris stream). Such structures were seen to be produced in disk-star collisions in a recent paper simulating the collision process \citep{Yao24}, but the argument put forward in this paper is independent of this finding, and is based only on energetic grounds and mass and angular momentum conservation.  These larger debris structures mean that the timescale for the flare is no longer set by photon diffusion (instead by the vertical extent of the debris), and that it is possible  that enough mass can be swept up in a collision to power the observed energies of long-period QPEs (although the we stress this must be tested with simulations, as the observations appear to be at the upper limit of what is possible to power by collisions). 

The focus of this paper is exclusively about studying the energetics, luminosity, temperatures and duration of the flares predicted in these TDE disk-collision models. We do not consider broader timing features (such as the observed ``long-short'' behavior of some systems \citealt{Miniutti2019}, or more erratic timing features of other sources, e.g., \citealt{Chakraborty25}, or flares disappearing and then reappearing again \citealt{Miniutti2023}). The reason for this is we believe that ensuring these models can reproduce the energetics of the events should be the first concern (for if they cannot one does not need to worry about timing features), and timing questions can be posed later. The physical properties of the extended structures which need to be invoked to explain the energetics of these events (within the collision paradigm), may indeed produce interesting timing features, which should be considered carefully.

The layout of this Paper is as follows. In section \ref{theory} we begin by discussing the differences between steady-state AGN-like disk models and those disks which form in the aftermath of a TDE which are highly constrained by mass and angular momentum conservation and the properties of the incoming star. We then derive a quiescent disk theory which takes into account the finite mass and angular momentum budget available to a TDE disk. We then re-perform the analysis of \cite{Linial2023}, with this modified quiescent disk model and extensions to different non-stellar-surface collisions, deriving a set of observational predictions. These observational predictions are now in fact over constrained, with 3 observable properties (luminosity, temperature and timescale) set by only two physical quantities (assuming the black hole mass is known): the effective area of the object colliding with the disk, and a product of quantities relating to the star which was tidally disrupted (this product will be defined precisely below). This means plausibility checks of the model can be performed, for any QPE unambiguously detected after a TDE, as this analysis removes unobservable nuisance parameters like the disk turbulence $\alpha$-parameter from the problem. In section \ref{obs} we do this plausibility test for AT2019qiz, finding the only remaining parameter space not ruled out by the QPE observables is that of an extended debris structure.  We also discuss the broader population of QPE sources.  In section \ref{conc} we conclude. Some technical results are presented in Appendices \ref{app}, \ref{app:height} and \ref{app:evolve}. 

\section{EMRI-TDE disk collisions }\label{theory}
\subsection{The mass in AGN-like steady state disks}
We begin with the standard steady-state \cite{SS73} thin disk accretion solutions for the disk
surface density  in a radiation pressure dominated flow, namely 
% \begin{equation}
%     {h \over r} \approx {3 \over 2 \epsilon } {r_g \over r} {\dot M \over \dot M_{\rm edd}}, 
% \end{equation}
% and 
\begin{equation}\label{eq:densAGN}
    \Sigma^{\rm SS}_{\rm rad} \approx {\dot M \over 3 \pi \nu} = {\dot M \over 3 \pi (GM_\bullet r)^{1/2} \alpha (h/r)^2}  \simeq 1.8 \times 10^{4} \, {\rm g/cm}^2 \, \left({0.1 \over \alpha}\right) \left({0.1 \dot M_{\rm edd} \over \dot M}\right) \left({r \over 100r_g}\right)^{3/2} , 
\end{equation}
which forms the basis of the disk properties in the detailed collisional QPE models \citep[e.g.,][]{Linial2023}. In these expressions symbols have their usual meanings: $\epsilon$ is the radiative efficiency of the flow, $\dot M$ is a radius-independent mass accretion rate through the disk, $\dot M_{\rm edd}\equiv L_{\rm edd} / \epsilon c^2$ is the Eddington accretion rate, $\alpha$ is the usual \cite{SS73} viscosity ($\nu$) parameter, and $r_g \equiv GM_\bullet/c^2$ is the gravitational radius of the black hole (which has mass $M_\bullet$). 

If, instead, we were to assume a gas-pressure dominated \cite{SS73} model (maybe more plausible for a large radius, long period, QPEs in the collision model framework), then we would find 
\begin{equation}\label{eq:densAGN_g}
    \Sigma^{\rm SS}_{\rm gas} \simeq 7.8 \times 10^{5} \, {\rm g/cm}^2 \, \left({0.1 \over \alpha}\right)^{4/5} \left({\dot M \over 0.1 \dot M_{\rm edd} }\right)^{3/5} \left({M_\bullet \over 10^6 M_\odot}\right)^{1/5} \left({r \over 100r_g}\right)^{-3/5} . 
\end{equation}
In either case, in the disk-EMRI models there is a second object which collides with the disk twice per orbital period. These collisions are assumed to be the origin of the QPEs, and the QPE period is given by one half of the EMRI's Keplerian period\footnote{We neglect orbital eccentricity throughout this paper for simplicity. Extending these results to include $e \neq 0$ would not be especially difficult, and will change the results presented here by order unity factors. Insight into the effects of eccentricity can be found in \citealt{Linial2023}. }  
\begin{equation}
    P_{\rm orb}  = 2 P_{\rm QPE} = 2\pi \sqrt{R_{\rm QPE}^3\over GM_\bullet} .
\end{equation}
It will be helpful to write 
\begin{equation}
    R_{\rm QPE} = \left({P_{\rm QPE}\over \pi}\right)^{2/3} (GM_\bullet)^{1/3}. 
\end{equation}
Combining the surface density implied from equation \ref{eq:densAGN} or \ref{eq:densAGN_g} with the definition of the QPE radius, the mass in the disk interior to the QPE radius is simply given by\footnote{We stress that this distinction is important even for the relatively small disks which fit within a $P_{\rm orb} \sim 40-72$ hour orbit for a $M_\bullet \sim 10^6 M_\odot$ black hole. Indeed, the values of $M_\bullet, \dot M$ and $P_{\rm QPE}$ inferred for AT2019qiz would imply in a radiation pressure dominated disk that $M^{\rm AT2019qiz}_{\rm disk}\approx 200 M_\odot \left({0.01 /\alpha}\right)$, while AT2022upj would have $M^{\rm AT2022upj}_{\rm disk} \approx 150 M_\odot \left({0.01 /\alpha}\right)$, clearly incompatible with them being TDE disks. }
\begin{align}
    M^{\rm SS}_{\rm disk}(r< R_{\rm QPE}) &= \int^{R_{\rm QPE}}_0 2\pi r' \Sigma(r') \, {\rm d}r' ,\\
    &\propto R_{\rm QPE}^{7/2} \propto P_{\rm QPE}^{7/3}, \quad {\rm (radiation \, pressure\, dominated)}, \\
    &\propto R_{\rm QPE}^{7/5} \propto P_{\rm QPE}^{14/15}, \quad {\rm (gas \, pressure\, dominated)}. 
\end{align}
We draw attention to a crucial point at this stage in the analysis, namely that steady-state, AGN-like disks are {\it significantly more massive the further away you are from the central black hole}, regardless of how one models the transport in the disk. This is because steady-state accretion disk models implicitly assume an infinite mass supply available to the disk. 

While a growing mass with radius is of course exactly how a steady-state AGN accretion flow will behave, the exact opposite behavior will be observed in  a TDE disk. The reason for this is two-fold. Firstly, the total mass content of a TDE disk has an absolute upper bound set by the mass of the  disrupted star, but more relevantly the mass in a disk (of given size) is more strongly limited by the {\it angular momentum} of the incoming stars orbit. 

\subsection{The absolute maximum mass of a TDE disk of fixed size}
Accretion disks are rotationally supported fluid systems, which means that if one knows two out of the (i) total angular momentum, (ii) total mass, or (iii) the size of the disk, one can constrain the third. For a TDE, we know {\it apriori} bounds on both the angular momentum and mass of the disk, as they are set by the properties of the incoming star. This allows us to use the size of the disk as a constraint on its remaining mass.  

The reason that angular momentum conservation places a strong limit on the total mass of the disk  is that, while mass might be harder to keep track off in a TDE (mass may be launched in outflows, scattered onto strange orbits, or take a very long time to settle into a disk), angular momentum is strictly conserved (as there is no process which can generate angular momentum in the disruption or disk-formation process) and importantly scales with $\sim r^{1/2}$ meaning that one cannot move material too far out, while conserving angular momentum, without reducing the mass in the disk. 

The angular momentum of the incoming star (and therefore the maximum angular momentum available to the disk) is well constrained for a TDE, as we know that the star must have been on a parabolic\footnote{If the orbit is not strictly parabolic this only changes the argument by a very small factor, and cannot change the scaling relationships derived here.} orbit which crossed into its tidal radius, and so
\begin{equation}
    J^{\rm tde}_\star = M_\star^{\rm tde} \sqrt{2 GM_\bullet r^{\rm tde}_T/\beta} .
\end{equation}
Where here we have introduced the stars mass $M_\star^{\rm tde}$ and radius $R_\star^{\rm tde}$, which has entered its tidal radius, which is roughly given by 
\begin{equation}
    r^{\rm tde}_T \approx R_\star^{\rm tde} \left({ M_\bullet \over M_\star^{\rm tde} }\right)^{1/3} ,
\end{equation}
and we have introduced the orbital penetration factor $\beta \equiv r_p / r_T \geq 1$, where $r_p$ is the pericenter of the orbit of the star (which sets its angular momentum).

Even if we were to conspire, by hand, to construct a disk with outer edge at $r_{\rm out} = R_{\rm QPE}$ with all of the mass and angular momentum of the initial star, the disk must still satisfy the twin constraints 
\begin{align}
    M_{\rm disk} &= \int^{R_{\rm QPE}}_0 2\pi r' \Sigma(r') \, {\rm d}r' \leq M_\star^{\rm tde} , \\
     J_{\rm disk} &= \int^{R_{\rm QPE}}_0 2\pi r' \sqrt{GM_\bullet r'} \, \Sigma(r') \, {\rm d}r' \leq J^{\rm tde}_\star. 
\end{align}
If we assume that the surface density of the disk is a constant with radius\footnote{This assumption introduces only order 1 constants into the angular momentum conservation limit, and does not change the scaling, see Appendix \ref{app}.}, then we find that the absolute maximum mass of the disk we can construct is 
\begin{equation}
    M_{\rm disk, \, MAX}^{\rm QPE} = {\rm minimum\, of\,} \begin{cases}
         M_\star^{\rm tde} \left({25 r^{\rm tde}_T / 8 \beta R_{\rm QPE}}\right)^{1/2}, \quad {\rm (angular\, momentum\, conservation)}, \\ \\
        M_\star^{\rm tde} , \quad {\rm (mass\, conservation)},
    \end{cases}
\end{equation}
where the upper limit is from angular momentum conservation, and the lower from mass conservation (see Appendix \ref{app} for a derivation). As these are dual constraints, the lower value must be taken of the two. 

The angular momentum limit is the relevant limiting factor if 
\begin{equation}
    R_{\rm QPE} > {25 r_T^{\rm tde} \over 8\beta}, 
\end{equation}
which assuming that the orbiting body in the problem would be disrupted if it approached $\sim r^{\rm tde}_T$ (i.e., if we assume that the object on the EMRI has similar density to the object which was disrupted), is likely to be the more relevant one. 

Steady-state models for accretion flows will violate these conservation laws for even moderate $r_{\rm out} \sim {\cal O}(10{\rm 's})\,  r_g$ disks, as can be seen both by their mass contents constructed earlier, and also by calculating the total angular momentum of these steady state disks 
\begin{align}
    J^{\rm SS}_{\rm disk}(r< R_{\rm QPE}) &= \int^{R_{\rm QPE}}_0 2\pi r' \sqrt{GM_\bullet r'} \, \Sigma(r') \, {\rm d}r', \\
    &\propto R_{\rm QPE}^{4} \propto P_{\rm QPE}^{8/3}, \quad {\rm (radiation \, pressure\, dominated)}, \\
    &\propto R_{\rm QPE}^{19/10} \propto P_{\rm QPE}^{19/15}, \quad {\rm (gas \, pressure\, dominated)}. 
\end{align}
This result (the upper bound on the disk mass) has serious implications for models which invoke disk-orbiter collisions as the origin of QPEs. For larger period systems (like AT2019qiz and AT2022upj), steady state models assume that the mass in the disk grows strongly ($\propto P_{\rm QPE}^{7/3}$, or $\propto P_{\rm QPE}^{19/15}$ for a gas pressure dominated disk), meaning that a growing energy budget is available in the collision for larger QPE period systems. However, as a TDE disk has a {\it fixed mass budget} there will in reality be {\it less} energy available in long QPE period collisions. Before we do any detailed computations, it can already be seen that this will be  problematic for the collisional QPE paradigm, as the natural scale of the mass hit during a collision will be 
\begin{equation}
    M_{\rm hit} \sim M_{\rm disk} \left({R_{\rm hitting} \over R_{\rm QPE}}\right)^2 \sim A_{\rm collide}R_{\rm QPE}^{-5/2} \sim A_{\rm collide} P_{\rm QPE}^{-5/3}, 
\end{equation}
where $R_{\rm hitting}$ is the effective size of the object doing the hitting, and the final scaling uses the angular momentum limited disk mass from above. The energy in the collision is then 
\begin{equation}
    E_{\rm hit} \sim {1\over 2} M_{\rm hit} v_{\rm hit}^2 \sim  M_{\rm disk} \left({R_{\rm hitting} \over R_{\rm QPE}}\right)^2 {GM_\bullet \over R_{\rm QPE}} \sim A_{\rm collide}R_{\rm QPE}^{-7/2} \sim A_{\rm collide} P_{\rm QPE}^{-7/3} . 
\end{equation}
This scaling of $E_{\rm hit} \sim P_{\rm QPE}^{-7/3}$ is, in strong contention with the data (Figure \ref{fig:data}). This suggests that the colliding area $A_{\rm collide}$ and the efficiency of turning this kinetic energy into radiation $\lambda$ must, if collisions are to explain QPE observations, scale strongly with $P_{\rm QPE}$\footnote{This may initially sound highly contrived, but will actually be the case in any of the following three plausible limits, (i) where the orbiting object is a black hole (with the collision size therefore set by its Bondi radius),  (ii) the stellar EMRI has ``puffed up'' due to repeated collisions with the disk to fill its Hills sphere, or (iii) the collisional is predominantly between a stream of stripped stellar debris and the disk. In the two former cases $R_{\rm hitting} \sim R_{\rm QPE} \sim P_{\rm QPE}^{2/3}$ and therefore $E_{\rm hit} \sim P_{\rm QPE}^{-1}$. All three of these situations will be dealt with in later sections of the paper.}. This immediately rules out a colliding stellar surface $A_{\rm collide} \sim \pi R_{\star}^2$ (i.e., an area which does not know about the orbital period) as a model of QPEs in TDE disks, as the radiative efficiency growth required to reproduce the observations $\lambda \sim (P_{\rm QPE})^{n}, n \gtrsim 3,$ would be highly fine tuned. 

To understand the properties of flares induced by collisions between an orbiting EMRI and a TDE disk, one must construct a quiescent disk theory which explicitly includes the constraints of TDE physics. As we shall demonstrate, this removes nuisance factors like $\alpha$ and $\dot M$ from the problem, and observables are more readily relatable to the properties of the two objects assumed to be in the system (the one disrupted and the one on an EMRI). This therefore facilitates plausibility tests of these models.

\subsection{Quiescent TDE disk theory}
In this section we move beyond the absolute maximum mass that can be in a TDE disk, to a more realistic description of the disk which is likely to be formed in such a system.  As discussed above, a tidal disruption event occurs when a star of mass $M_\star^{\rm tde}$ and radius $R_\star^{\rm tde}$ enters its tidal radius, which is roughly given by 
\begin{equation}
    r^{\rm tde}_T \approx R_\star^{\rm tde} \left({ M_\bullet \over M_\star^{\rm tde} }\right)^{1/3} ,
\end{equation}
after which half of the star remains bound, and half is ejected on hyperbolic orbits \citep{Rees88}. An initially complex series of hydrodynamic process ultimately results in the circularisation of a total fraction $0 < f_d < 1$ of this bound material, at roughly the circularisation radius $r^{\rm tde}_c = 2r^{\rm tde}_T/\beta$, where the factor two results from angular momentum conservation during the circularisation process. 

The initial mass in the disk we shall define by 
\begin{equation}
    M_{\rm disk, 0} = {1\over 2}f_d M_\star^{\rm tde} ,
\end{equation}
which incorporates both the factor $1/2$ of the mass which is unbound, and the additional nuisance parameter $f_d$ which allows for inefficient disk formation. This mass then begins to be accreted and drops with time. The total angular momentum of the disk is conserved during this process \citep[see e.g.,][]{LBP74, Cannizzo90}, meaning that the outer edge of the disk must grow with time to compensate for the dropping mass of the flow, as the quantity 
\begin{equation}
    J_{\rm disk}(t) \propto M_{\rm disk} (t) \sqrt{R_{\rm out}(t)} \simeq {\rm constant}. 
\end{equation}
This means that by the time the disk spreads to $R_{\rm out}(t) \simeq R_{\rm QPE}$, the disk has a surviving mass content 
\begin{equation}
    M_{\rm disk}^{\rm QPE} = {1\over 2} f_d M_\star^{\rm tde} \left({2r^{\rm tde}_T \over \beta R_{\rm QPE}}\right)^{1/2}   \approx 0.5 M_\odot  \, {f_d \over \beta^{1/2}}  \left({M_\star^{\rm tde}\over M_\odot}\right)^{5/6} \left({R_\star^{\rm tde} \over R_{\odot}}\right)^{1/2} \left({4\, {\rm hours} \over P_{\rm QPE}}\right)^{1/3} .%{f_d\over\sqrt{2\beta}} M_\star^{\rm tde} \left({P_\star^{\rm tde} \over P_{\rm QPE}}\right)^{1/3} \approx 0.5 M_\odot  \, {f_d \over \beta^{1/2}}  \left({M_\star^{\rm tde}\over M_\odot}\right)^{5/6} \left({R_\star^{\rm tde} \over R_{\odot}}\right)^{1/2} \left({4\, {\rm hours} \over P_{\rm QPE}}\right)^{1/3},
\end{equation}
%where we have defined a dynamical time of the star which was tidally disrupted 
%\begin{equation}
%    P_\star^{\rm tde} \equiv \pi \sqrt{(R_\star^{\rm tde})^3 \over GM_\star^{\rm tde}} = 1.4\, {\rm hours}\, \left({M_\star^{\rm tde}\over M_\odot}\right)^{-1/2} \left({R_\star^{\rm emri} \over R_{\odot}}\right)^{3/2} ,
%\end{equation}
Note that this is independent of black hole properties, and is in fact equal to a factor $2f_d/5$ times the maximum possible value allowed by angular momentum conservation. We have implicitly assumed in this calculation that $R_{\rm QPE} > 2r_T/\beta$, which amounts to assuming that the object on the EMRI is not significantly more dense than the star which was disrupted (and is therefore not orbiting inside of the tidal radius of the TDE star). If $R_{\rm QPE} < 2r_T/\beta$, then the total mass in the disk should be taken to be $f_d M^{\rm tde}_\star/2$. We do not explicitly carry these if/else statements through the rest of this paper, but these checks are implemented in the numerical model used later. 

The height-integrated surface density of the disk at the location at which an object with orbital period $P_{\rm orb} = 2 P_{\rm QPE}$ would intercept with it is\footnote{This assumes a constant surface density as a function of radius in the disk. This assumption does not effect the results presented here, other than through order unity constants, see Appendix \ref{app}.  } 
\begin{equation}
\Sigma^{\rm QPE}_{\rm disk, TDE} \approx {M_{\rm disk}^{\rm QPE} \over \pi R_{\rm QPE}^2} , 
\end{equation}
or explicitly 
\begin{equation}\label{densTDE}
\Sigma^{\rm QPE}_{\rm disk, TDE} \approx 1.6\times10^6 \, {\rm g\, cm}^{-2} {f_d \over \beta^{1/2}} \left({M_\star^{\rm tde}\over M_\odot}\right)^{5/6}   \left({R_\star^{\rm tde} \over R_{\odot}}\right)^{1/2} \left({4 \, {\rm hours} \over P_{\rm QPE}}\right)^{5/3} \left({10^6 M_\odot \over M_\bullet}\right)^{2/3} .
\end{equation}
If we contrast this with the quiescent disk model of \cite{SS73} which when expressed in these units is 
\begin{equation}
\Sigma^{\rm SS}_{\rm disk, rad} \approx 1.8\times10^4 \, {\rm g/cm}^2 \, \left({0.1 \over \alpha}\right) \left({0.1 \dot M_{\rm edd} \over \dot M}\right)  \left({P_{\rm QPE} \over 4\, {\rm hours}}\right) \left({10^6 M_\odot \over M_\bullet}\right) ,
\end{equation}
we see that a TDE disk is much more dense for short-period QPEs than previously anticipated, with a QPE-period dependence which points in the opposite direction. This modifies all of the predictions of the theory, as we now elucidate. In effect, very little of the analysis in the next subsection is new, and is just a re-running of the \cite{Linial2023} analysis with a modified disk surface density profile. The reader not so interested in the technical details may wish to skip to section \ref{obs}, where these models are compared to observations.

\subsection{TDE disk-orbiter collision theory -- ``hard sphere'' limit}
Assuming that this mass is spread uniformly within the disk, then the amount of material which will be ejected in the collision between this disk and an orbiting body is 
\begin{equation}
    M_{\rm ej} \approx 2\pi (R_\star^{\rm emri})^2 \, \Sigma_{\rm disk} \approx 2 \left({R_\star^{\rm emri} \over R_{\rm QPE}}\right)^2 M_{\rm disk}^{\rm QPE} , 
\end{equation}
where the factor 2 results from the fact that both the disk and EMRI star are orbiting at the same velocity, but in roughly perpendicular directions (and so the disk rotates into the orbiting body as it passes through). 

In this first subsection we shall follow the original calculations of \cite{Linial2023} in assuming that the orbiting body is a star (hence the notation $R_\star^{\rm emri}$), but this of course does not necessarily mean that the intercepting surface area is just the area of the stellar surface (a situation which hereafter we shall refer to as the ``hard sphere'' EMRI limit). The star which undergoes collisions may be ``puffed up'' by these collisions, increasing its surface area out to its Hills sphere, or the collision could be between a stream of stellar debris (from previous collisions) and the disk \citep[as in the numerical findings of][]{Yao24}. Alternatively, this orbiting cross section could be the Bondi radius of a black hole. All of these cases will be examined in detail in further sections. 

In the ``hard sphere'' EMRI limit, this ejected mass evaluates to 
\begin{equation}
M_{\rm ej} \approx 2.4 \times 10^{-5} \, M_\odot \, {f_d \over \beta^{1/2}} \left({M_\star^{\rm tde}\over M_\odot}\right)^{5/6}  \left({R_\star^{\rm tde} \over R_{\odot}}\right)^{1/2}  \left({R_\star^{\rm emri} \over R_{\odot}}\right)^2 \left({4 \, {\rm hours} \over P_{\rm QPE}}\right)^{5/3} \left({10^6 M_\odot \over M_\bullet}\right)^{2/3} . 
\end{equation}
We can therefore constrain an absolute maximum energy budget available to the collision which is $E_{\rm ej} =  v_{K}^2 M_{\rm ej}$, or 
\begin{equation}
    E_{\rm ej} \approx {GM_\bullet \over R_{\rm QPE}} M_{\rm ej},%=  \sqrt{2} \pi^2 {f_d \over \beta^{1/2}} M_\star^{\rm tde} \left({R_\star^{\rm emri} \over P_{\rm QPE}}\right)^2 \left({P_\star^{\rm tde} \over P_{\rm QPE}}\right)^{1/3} . 
\end{equation}
substituting in the above scalings we find the elegant result that this energy scale only depends on the properties of the two stars $R_\star^{\rm emri}, R_\star^{\rm tde}, M_\star^{\rm tde}$ and the observable $P_{\rm QPE}$. The scaling is as follows 
\begin{equation}
    E_{\rm ej} \approx 4.6 \times 10^{47} \, {\rm erg} \,  {f_d\over \beta^{1/2}} \left({M_\star^{\rm tde}\over M_\odot}\right)^{5/6} \left({R_\star^{\rm tde} \over R_{\odot}}\right)^{1/2} \left({R_\star^{\rm emri} \over R_{\odot}}\right)^2 \left({4 \, {\rm hours} \over P_{\rm QPE}}\right)^{7/3}. 
\end{equation}
Note the rapid drop in the available energy budget for long period QPEs $E_{\rm ej} \propto P_{\rm QPE}^{-7/3}$, and the lack of black hole parameter dependence. The ejected mass also sets the timescale for which the QPE flare lasts, as it sets the optical depth of the shocked gas as it expands 
\begin{equation}
    \tau(t) = {\kappa_{\rm es} M_{\rm ej} \over 4\pi R_{\rm ej}^2} ,
\end{equation}
and we will assume that the ejecta expand on ballistic trajectories $R_{\rm ej} \propto v_K t$ for times $t \gg R_\star^{\rm emri}/v_K$ (the initial optical depth is just equal to that of the disk, i.e., $\tau(0) = \tau_{\rm disk} \gg 1$, and therefore the timescale for the optical depth to drop comfortably satisfies this constraint). Treating the subsequent evolution of this system just like a supernovae shock breakout, observable emission escapes when the optical depth reaches $c/v_K$\footnote{If we made the assumption that photons only escape when the optical depth of the ejecta drops to $\tau \approx 1$, then subsequent scaling relationships are modified by another multiplicative factor of $t_{\rm diff} \to t_{\rm diff}(c/v_K)^{1/2}$ which changes the scalings by a factor $\sim  P_{\rm QPE}^{1/3} M_\bullet^{1/6}$.}, meaning that the time taken for the optical depth of the cloud to drop sufficiently is, up to order unity constants, equal to 
\begin{equation}
    t_{\rm diff} \approx \left({\kappa_{\rm es} M_{\rm ej} \over 4 \pi c v_K}\right)^{1/2} . %= \sqrt{{\kappa_{\rm es} f_d M_\star^{\rm tde}  \over 4 \pi \beta^{1/2} c v_\star^{\rm emri}} } \left({P_\star^{\rm emri} \over P_{\rm QPE}}\right)^{1/2} \left({P_\star^{\rm tde} \over P_{\rm QPE}}\right)^{1/6} \left({M_\star^{\rm emri} \over M_\bullet}\right)^{1/2} , 
\end{equation}
%where we have introduced the characteristic dynamical velocity of the EMRI star 
%\begin{equation}
%    v_\star^{\rm emri} \equiv \sqrt{GM_\star^{\rm emri} \over R_\star^{\rm emri}} .
%\end{equation}
Evaluating this we find 
\begin{equation}\label{tdiffsphere}
    t_{\rm diff} \approx 1.1\, {\rm hours }\,  {f_d^{1/2} \over \beta^{1/4}} \left({M_\star^{\rm tde}\over M_\odot}\right)^{5/12} \left({R_\star^{\rm tde} \over R_{\odot}}\right)^{1/4}   \left({R_\star^{\rm emri} \over R_{\odot}}\right)  \left({4 \, {\rm hours} \over P_{\rm QPE}}\right)^{2/3} \left({10^6 M_\odot \over M_\bullet}\right)^{1/2} .
\end{equation}
In the hard sphere limit we have that $t_{\rm diff } \gg t_{\rm cross} \sim \max(h, R_\star^{\rm emri})/v_K$ (where $h$ is the height of the disk; note that the opposite limit $t_{\rm cross} \gg t_{\rm diff}$ is something which is possible for collisions between a  long debris stream and a disk, as we shall discuss later). This implies that the time of the QPE flare $t_{\rm QPE}\sim t_{\rm diff}$, and that hard-sphere QPE's have an associated duty cycle of 
\begin{equation}
    {\cal D} \equiv {t_{\rm QPE} \over P_{\rm QPE}} \approx 0.27  \, {f_d^{1/2}\over \beta^{1/4}} \left({M_\star^{\rm tde}\over M_\odot}\right)^{5/12}  \left({R_\star^{\rm tde} \over R_{\odot}}\right)^{1/4}   \left({R_\star^{\rm emri} \over R_{\odot}}\right)\left({4 \, {\rm hours} \over P_{\rm QPE}}\right)^{5/3} \left({10^6 M_\odot \over M_\bullet}\right)^{1/2} . 
\end{equation}
This scaling ${\cal D}\sim P_{\rm QPE}^{-5/3}$ is in strong contention with the $\sim$ constant duty cycle observed in QPE sources. 

The ejecta resulting from this collision is initially optically thick and rapidly moving, meaning that the size of the ejecta has expanded significantly by the time emission escapes the cloud. Therefore the initial kinetic energy in the cloud undergoes adiabatic losses, reducing the energy budget that can be radiated away. The initial density of the post-shock cloud is $\rho_{\rm sh} \approx \rho_{\rm disk} (\gamma + 1)/(\gamma - 1)  = 7 \rho_{\rm disk}$, where $\rho_{\rm disk}$ is the density of the gas in the disk (this is just a statement of the strong shock conditions which are justified owing to the supersonic velocity of the orbiting object $v_K/c_s^2 \sim (h/r)^{-2} \gg 1$), and we assume a radiation pressure dominated equation of state ($\gamma = 4/3$).  This means that the width of the ejecta is initially $w \approx h/7$ where $h$ is the scale height of the disk.  The size of the ejecta once it begins to emit photons is $R_{\rm ej} \approx v_K t_{\rm diff}$, and so the energy available to be radiated from the cloud is $E_{\rm rad} \approx E_{\rm ej} (3V_0/4\pi R_{\rm ej}^3)^{\gamma - 1} $, where $V_0\approx (R_\star^{\rm emri})^2 w$, or explicitly 
\begin{equation}
E_{\rm rad} \approx E_{\rm ej}  \left({3 (R_\star^{\rm emri})^2 h \over 28  v_K^3 t_{\rm diff}^3 }\right)^{1/3} , 
\end{equation}
which once simplified %is 
%\begin{equation}
%E_{\rm rad} \approx E_{\rm ej}  \left({3 \over 28}\right)^{1/3} {1\over \pi} \left({h\over r}\right)^{1/3} \left({P_\star^{\rm emri} \over t_{\rm QPE}}\right)^{4/9}  \left({P_{\rm QPE} \over t_{\rm QPE}}\right)^{5/9} \left({M_\star^{\rm emri} \over M_\bullet}\right)^{2/9} ,
%\end{equation}
%or 
is explicitly 
\begin{equation}\label{Eradsphere}
E_{\rm rad} \approx 3.5 \times 10^{45} \, {\rm erg} \, {f_d^{1/2}\over \beta^{1/4}} \left({M_\star^{\rm tde}\over M_\odot}\right)^{5/12} \left({R_\star^{\rm tde} \over R_{\odot}}\right)^{1/4}\, \left(10{h\over r}\right)^{1/3}  
\left({R_\star^{\rm emri} \over R_{\odot}}\right)^{5/3} \left({4 \, {\rm hours} \over P_{\rm QPE}}\right)^{10/9} \left({ M_\bullet \over 10^6 M_\odot}\right)^{1/8} . 
\end{equation}

We see that at this point we have our first ``unknown'' from the disk, namely the local disk scale height at the collision radius $h/r$\footnote{We show in Appendix \ref{app:height} that this is not really an unknown if the temperature of the inner disk is constrained from observations, which it often is for a given QPE source. We however treat this as an unknown for the remainder of the manuscript, for simplicity.}. Fortunately this quantity enters only as weak powers throughout the analysis. This means that the efficiency at which kinetic energy from the collision is turned into radiation is
\begin{equation}
    \lambda \equiv {E_{\rm rad}\over E_{\rm ej}} = 7.8\times 10^{-3} {f_d^{-1/2}\over \beta^{-1/4}} \left({M_\star^{\rm tde}\over M_\odot}\right)^{-5/12} \left({R_\star^{\rm tde} \over R_{\odot}}\right)^{-1/4}\, \left(10{h\over r}\right)^{1/3}  
\left({R_\star^{\rm emri} \over R_{\odot}}\right)^{-1/3} \left({P_{\rm QPE} \over 4 \, {\rm hours} }\right)^{11/9} \left({ M_\bullet \over 10^6 M_\odot}\right)^{-1/8} 
\end{equation}

As the emission escapes over a timescale $t_{\rm diff}$, we can evaluate a luminosity scale of QPE emission 
\begin{equation}
L_{\rm flare} \approx {E_{\rm rad} \over t_{\rm diff}},  %\approx {E_{\rm ej} \over t_{\rm QPE}} \left({3 \over 28}\right)^{1/3} {1\over \pi} \left({h\over r}\right)^{1/3} \left({P_\star^{\rm emri} \over t_{\rm QPE}}\right)^{4/9}  \left({P_{\rm QPE} \over t_{\rm QPE}}\right)^{5/9} \left({M_\star^{\rm emri} \over M_\bullet}\right)^{2/9} . 
\end{equation}
which when simplified is  
\begin{equation}
L_{\rm flare} \approx 9.4 \times 10^{41} \, {\rm erg\, s}^{-1} \,\left(10{h\over r}\right)^{1/3}  \left({R_\star^{\rm emri} \over R_{\odot}}\right)^{2/3}  \left({4 \, {\rm hours} \over P_{\rm QPE}}\right)^{4/9} \left({ M_\bullet \over 10^6 M_\odot}\right)^{5/8} . 
\end{equation}
Note that the properties of the disrupted star have completely dropped out of this expression. This is because the mass in the ejecta cancels, as first noted by \cite{Linial2023} (i.e., this is independent of disk surface density as, briefly, having more mass in the ejecta allows for a larger energy budget to be radiated, but also results in a larger initial optical depth meaning that this extra energy is radiated over a longer timescale, leading to a luminosity which is independent of ejecta mass). The effective blackbody temperature of this radiation is then 
\begin{equation}
k T_{\rm BB} =  k \left({c u_\gamma \over 4\sigma}\right)^{1/4} \approx k \left({3 c E_{\rm rad} \over 16 \pi \sigma R_{\rm ej}^3}\right)^{1/4}   = k \left({3 c E_{\rm rad} \over 16 \pi \sigma v_{K}^3 t_{\rm diff}^3}\right)^{1/4} , 
\end{equation}
where $u_\gamma$ is the radiation energy density, and $\sigma$ is the Stefan-Boltzmann constant. This evaluates to 
\begin{equation}
k T_{\rm BB} \approx 7.9 \, {\rm eV} \, {\beta^{1/8} \over f_d^{1/4}} \, \, \left({M_\star^{\rm tde}\over M_\odot}\right)^{-5/24} \left({R_\star^{\rm tde} \over R_{\odot}}\right)^{-1/8} 
\left(10{h\over r}\right)^{1/12} \left({P_{\rm QPE} \over 4\, {\rm hours}}\right)^{17/36} \left({R_\star^{\rm emri} \over R_{\odot}}\right)^{-1/3}  \left({ M_\bullet \over 10^6 M_\odot }\right)^{-1/8} . 
\end{equation}
This is too soft an emission component to explain the observations $k T_{\rm obs} \approx 100-200$ eV. In the original TDE disk-EMRI model \citep{Linial2023}, the resolution of this discrepancy was believed to be because of a lack of thermal equilibrium in the expanding gas, and insufficient photon production, which hardened the resulting emission. Following this same argument again, we compute the dimensionless photon production efficiency coefficient \citep{Nakar10}
\begin{equation}
\eta \equiv {n_{\rm BB} (T_{\rm sh}) \over t_{\rm cross} \, \dot n_{{\rm ff}, \gamma} (\rho_{\rm sh}, T_{\rm sh})} . % \approx {280\, {\rm s} \over t_{\rm cross}} \left({\rho_{\rm sh} \over 10^{-10} \, {\rm g}/{\rm cm}^3}\right)^{-2} \, \left({k T_{\rm sh} \over 10 \, {\rm eV}}\right)^{7/2} . 
\end{equation}
Here $\dot n_{{\rm ff}, \gamma} \propto \rho^{2}T^{-1/2}$ is the photon production rate by free-free Bremsstrahlung, where the post shock density of the gas is gvien by $\rho_{\rm sh} = 7\Sigma_{\rm disk}/(2h)$,  $T_{\rm sh} = (3c \rho_{\rm sh}v_K^2 / 4\sigma)^{1/4}$ is the effective blackbody temperature of this shocked gas, and $t_{\rm cross} = h/(7v_K)$ is the time it takes for the object on an EMRI to cross the shocked gas. The denominator therefore represents the number of photons which can be produced during the period of highest density post shock. The numerator $n_{\rm BB}\approx 4\sigma T^3 / 3kc$ represents the number of photons which would have to be produced by Bremsstrahlung for the gas to reach thermodynamic equilibrium, meaning $\eta>1$ corresponds to photon starvation. Evaluating this criteria we find 
\begin{equation}
\eta \approx 0.02 \, {\beta^{9/16} \over f_d^{9/8}} \left(10{h\over r}\right)^{1/8} \, \left({ P_{\rm QPE} \over 4 \, {\rm hours} }\right)^{25/24} \, \left({M_\odot \over M_\star^{\rm tde}}\right)^{15/16} \,    \left({R_\odot \over R_\star^{\rm tde}}\right)^{9/16} \, \left({M_\bullet \over 10^6 M_\odot}\right)^{41/24} , 
\end{equation} 
meaning that for short period EMRI-TDE disk collisions, photon starvation is in general not important as sufficient photons are produced in the shocked gas to reach thermodynamic equilibrium. We note that this result is independent of the properties of the object on the EMRI, as this is simply a function of the thermodynamic properties of the post-shocked gas, and is not sensitive to what did the shocking \citep[see also][]{Linial2023}. 

For sufficiently long period QPEs, or high black hole masses, this is not true however, and the radiation will generally be released at harder energies than $kT_{\rm BB}$. As well as photon starvation, however, inverse Comptonization can be important in this regime. Inverse Comptonization becomes important when the Compton $y$ parameter exceeds unity $(y > 1)$. Following \cite{Nakar10}, we compute the Compton $y$ parameter using 
\begin{equation}
y = 3 \left({\rho_{\rm sh} \over 10^{-9} \, {\rm g}\, {\rm cm}^{-3}}\right)^{-1/2} \, \left({k T_{\rm sh} \over 100 \, {\rm eV}}\right)^{9/4} , 
\end{equation}
which evaluates to 
\begin{equation}
y \approx 7.6 \, {f_d^{1/16} \over \beta^{1/32} } \left(10 {h\over r}\right)^{-1/16} \left({M_\star^{\rm tde} \over M_\odot}\right)^{5/96} \left({R_\star^{\rm tde} \over R_{\odot}}\right)^{1/32}  \left({M_\bullet \over 10^6 M_\odot}\right)^{15/16} \left({P_{\rm QPE} \over 4\, {\rm hours} }\right)^{-25/48} , 
\end{equation}
meaning that in general inverse Comptonization is important for the gas shock heated in a TDE disk-EMRI collision (a result again independent of the properties of the object on an EMRI, and contrary to the findings of \citealt{Linial2023}, a result of the higher disk density in the short period TDE-QPE case). Inverse Comptonization limits the peak temperature reached by the photon starved gas through a logarithmic factor $\xi$
\begin{equation}
\xi = {1\over 2} \ln(y) \left[1.6 + \ln(y)\right] ,
\end{equation}
with for typical parameters is $\xi \gtrsim 3$. The observed temperature, when taking into account photon starvation and inverse Comptonization is \citep{Nakar10}
\begin{equation}
k T_{\rm obs} \approx k T_{\rm BB} \times 
\begin{cases}
&1, \quad \quad\quad \,\,\,  \eta < 1, \quad {\rm or} \quad   \xi > \eta > 1, \\
&\eta^2 , \quad\quad \quad  \eta > 1 \quad {\rm and} \quad  \xi < 1,\\
& \left({\eta /  \xi}\right)^{2} , \quad \eta >  \xi > 1.  
 \end{cases}
\end{equation} 
We find for typical parameters that this temperature is still too low to match QPE observations. 

\subsection{Material stripped from the EMRI by collisions cannot substantially raise the maximum density of a TDE disk}
Contrasting equations (\ref{Eradsphere}) and (\ref{tdiffsphere}) with the observed data (Figure \ref{fig:data}), we immediately see that collisions between a hard-sphere stellar EMRI and a TDE disk will be unable to reproduce either the amplitude or period-dependence of the energy or timescale of QPE flares. This, ultimately, is a result of TDE disks being significantly less dense than  steady-state AGN-like accretion flows at large radii.   Before we move on to different models for the collision geometry which may help resolve this tension, we note that it is not possible to resolve this energy discrepancy by invoking an increase in the {\it maximum} disk density during the TDEs evolution caused by material added to the disk from the EMRI. This can be seen without any reference to the process which strips material off the EMRI, and is a simple consequence of mass conservation. 

Imagine that an amount of material $\Delta M^{(1)}_{\rm emri}$ is stripped off the orbiting EMRI after each collision. The rate at which this mass is added to the disk $\dot M_{\rm add} \sim \Delta M^{(1)}_{\rm emri}/P_{\rm QPE}$ may well exceed the local accretion rate of the quiescent disk by a large fraction \citep[e.g.,][]{Linial2023}. In principle, one could imagine that the disk may settle into some longtime equilibrium, many years after the TDE, where the main mass content of the disk is set by this feeding. Independent of considerations about whether such a situation can occur, if one where to use this mass addition rate within a traditional $\alpha$-model framework then one would infer that the density of the disk would go up by a similarly large fraction (as in a steady state $\alpha$ model $\Sigma \propto \dot M$, this is just equation \ref{eq:densAGN} again). Such a conclusion is not 
correct, just as it was not for the original TDE disk, as it is not consistent with mass conservation. After $N$ collisions, when $\Delta M_{\rm emri} = N \Delta M^{(1)}_{\rm emri}$ is added to the disk, the most that the surface density can be increased by adding this material is 
\begin{equation}
    \Delta \Sigma_{\rm add} = {\Delta M_{\rm emri} \over 2\pi R_{\rm QPE} \Delta R}, 
\end{equation}
where $\Delta R$ is some radial scale which describes the width taken up by the debris which has been added to the disk (the material must be spread over the full $2\pi R_{\rm QPE}$ to have a realistic prospect of being involved in a future collision). The amount the disk surface density can be increased by collisions is then 
\begin{equation}
    {\Delta \Sigma_{\rm add} \over \Sigma_{\rm disk}} \sim {\Delta M_{\rm emri} \over M_{\rm disk}} {\Delta R \over R_{\rm QPE}} .
\end{equation}
Obviously $M^{\rm emri}_{\star}$ and $M_{\rm disk}$ have the same natural scale\footnote{Technically, EMRI-AGN disk collisions can substantially raise the surface density scale of the very small radius regions of steady-state AGN-like disks, as these very different disks can have mass contents which are $M_{\rm disk}^{\rm AGN}(r\sim 10r_g)\ll M_\odot$ in their innermost regions (i.e., equation \ref{eq:densAGN}). We note that the star on the EMRI would almost certainly have been destroyed at larger radii in the denser regions of the AGN disk before it reached such a scale, however. } (both are ultimately related to $\sim$ a solar masses worth of material). Even if all of the added matter were somehow contained within a width $\Delta R \sim R_\odot$ for a long period (something extremely unlikely to occur as any density gradients would lead to a quick evolution of the flow), the fact that $R_{\rm QPE} /R_\odot \sim 200 (P_{\rm QPE}/4{\rm hours})^{2/3} (M_\bullet/10^6 M_\odot)^{1/3}$, and that the radiated energy goes only as the square root of the disk surface density, implies that one cannot locally raise the surface density of the disk by the orders of magnitude required to explain the energies seen in long-period TDE-QPEs without both adding $\sim$ the entire EMRI star to the disk (thereby preventing any further collisions), and also somehow keeping it contained within a very narrow radial range. 

In reality of course the additional matter will spread all the way down to the inner edge of the disk, meaning that $\Delta R \sim R_{\rm QPE}$ and we find the extremely intuitive result 
\begin{equation}
    {\Delta \Sigma_{\rm add} \over \Sigma_{\rm disk}} \sim {\Delta M_{\rm emri} \over M_{\rm disk}} , \quad \Sigma_{\rm tot}  \approx {M_{\rm disk}^{\rm QPE} \over \pi R_{\rm QPE}^2}\left( 1 + {\Delta M_{\rm emri} \over M_{\rm disk}^{\rm QPE}}\right),
\end{equation}
i.e., the most you can increase the maximum surface density of a disk which was originally formed from a star by adding material from a second star is a factor $\sim 2$. In effect this is simply the same disk density limit which holds for the global TDE disk, but now for the amount of material added from the EMRI.

The argument in this section is related to attempting to raise the disk density above the maximum it reaches during the TDE evolution by adding material from the EMRI (this maximum density is the relevant density for comparison when worrying about the peak energetics of the flares). Of course, at very late times post TDE, it is possible that the disk density from the original stellar disruption will have decayed away to a very small value due to continued accretion (i.e., to a value much less than the value when the TDE disk first spread to $R_{\rm QPE}$), at which point the mass added from the  EMRI may change the disk density {\it at that late time} by a non-negligible fraction. This evolution will be complex, and we discuss it further in a later section, but it will likely in effect prevent the disk density from dropping below some minimum lower bound $\Sigma_{\rm disk}(t) \gtrsim \Sigma_{\rm min}$, which is much lower than the density reached during the peak of the TDE evolution $\Sigma_{\rm min} \ll \Sigma_{\rm disk, TDE}^{\rm QPE}$. The disk dynamics of this later time, low density, state are likely very interesting \citep{Linial24}.

As noted above, the  expressions of the previous section all assumed that the orbiting body acted as a hard sphere, which is not the only possible outcome \citep[in fact it is contrary to numerical simulations of the collision process][]{Yao24}. In the following four subsections we consider different collisional cross section possibilities.

\subsection{ An intermediate mass black hole as the EMRI object } 
The above arguments assumed that the orbiting object had a solid surface with which it could collide with the disk. If this is not the case, and the orbiting object is an intermediate mass black hole \citep[e.g.,][]{Franchini2023}, then the relevant radial scale is the Bondi radius 
\begin{equation}
R_B = {Gm_\bullet \over |\vec v_{\rm rel}|^2} \approx {Gm_\bullet \over v_K^2} f(\psi)  \approx 0.2 R_\odot \left({m_\bullet \over 10^3 M_\odot}\right) \left({M_\bullet \over 10^6 M_\odot}\right)^{-2/3}\, \left({P_{\rm QPE} \over 4\, {\rm hours}}\right)^{2/3} f(\psi) .
\end{equation}
The relative velocity between the disk and the IMBH is $\vec v_{\rm rel} = \vec v_\bullet - \vec v_{\rm disk}$, and thus $|\vec v_{\rm rel}|^2 = \vec v_\bullet^2 + \vec v_{\rm disk}^2 - 2 \vec v_\bullet \cdot \vec v_{\rm disk} \approx 2v_K^2 +  c_S^2 - 2 v_K^2 \sin \psi$, where $0<\psi<2\pi$ is the angle between the IMBH's orbital velocity and the central black hole's spin axis, and the speed of sound in the disk $c_S$ represents the random component of the velocity in the disk flow. We then define 
\begin{equation}
    f(\psi) \equiv {v_K^2 \over |\vec v_{\rm rel}|^2 } = {1 \over 2 - 2\sin \psi + (h/r)^2},
\end{equation}
where we have assumed that the disk is sufficiently thin that $(c_s^2/v_K^2 \sim (h/r)^2 \ll 1)$. We shall assume that the orbit of the IMBH is not fine tuned, such that $v_{\rm rel}^2\sim v_K^2$ (i.e., we take $f(\psi) = 1$ as default). If the IMBH is orbiting in the same direction as the disk  and at a very slight inclination from the disk plane $(\psi \sim \pi/2)$ then the orbital velocities will $\sim$ cancel and the Bondi radius will be enhanced by a factor $v_K^2/c_s^2\sim(r/h)^2$. Results in this extreme limit can be found by substituting $m_\bullet \to m_\bullet (r/h)^2$ throughout. This, however, cannot be invoked for more than a very small number of sources, as it is highly fine tuned.  Folding this Bondi radius through in place of  $R_\star^{\rm emri}$ we find the following scaling relationships for the four observable parameters 
\begin{equation}
    t_{\rm diff, \bullet} \approx 0.21\, {\rm hours }\,  {f_d^{1/2} \over \beta^{1/4}} \left({M_\star^{\rm tde}\over M_\odot}\right)^{5/12} \left({R_\star^{\rm tde} \over R_{\odot}}\right)^{1/4}   \left({10^6 M_\odot \over M_\bullet}\right)^{7/6} \left({m_\bullet \over 10^3 M_\odot}\right) \,f(\psi) ,
\end{equation}
\begin{equation}
L_{\rm flare, \bullet} \approx 3.2 \times 10^{41} \, {\rm erg\, s}^{-1} \,\left(10{h\over r}\right)^{1/3}  \left({ M_\bullet \over 10^6 M_\odot}\right)^{13/72} \left({m_\bullet \over 10^3 M_\odot}\right)^{2/3} \,[f(\psi)]^{2/3} , 
\end{equation}
\begin{equation}
E_{\rm rad, \bullet} \approx 2.5 \times 10^{44} \, {\rm erg} \, {f_d^{1/2} \over \beta^{1/4}} \left({M_\star^{\rm tde}\over M_\odot}\right)^{5/12} \left({R_\star^{\rm tde} \over R_{\odot}}\right)^{1/4}  \,\left(10{h\over r}\right)^{1/3}  \left({  10^6 M_\odot \over M_\bullet }\right)^{71/72} \left({m_\bullet \over 10^3 M_\odot}\right)^{5/3}\,[f(\psi)]^{5/3}  , 
\end{equation}
and 
\begin{equation}
k T_{\rm BB, \bullet} \approx 13.4 \, {\rm eV} \, {\beta^{1/8} \over f_d^{1/4}} \, \left(10{h\over r}\right)^{1/12} \left({P_{\rm QPE} \over 4\, {\rm hours}}\right)^{1/4} \,  \left({M_\star^{\rm tde}\over M_\odot}\right)^{-5/24} \left({R_\star^{\rm tde} \over R_{\odot}}\right)^{-1/8}  \left({ M_\bullet \over 10^6 M_\odot }\right)^{7/72} \left({m_\bullet \over 10^3 M_\odot}\right)^{-1/3}\,[f(\psi)]^{-1/3}  .
\end{equation}
The properties of the radiation field (i.e., the effects of photon starvation and inverse Comptonization) are unchanged in this scenario from that derived above, as they depend only on the properties of the disrupted star. As in \cite{Linial2023}, it does not appear that IMBH-disk collisions are luminous enough to adequately describe the observed QPE properties, except for $m_\bullet \gtrsim 10^5 M_\odot$. These large mass companions can be ruled out in one sense as the QPE rate is so high \citep[see][for a discussion of the volumetric rates of QPEs, and the rates of QPEs following TDEs]{Arcodia24b, Chakraborty25} that this many orbiting IMBH companions would grow the primary supermassive black holes too quickly (as they would inspiral due to gravitational wave emission), thereby preventing the detection of $\sim 10^6 M_\odot$ black holes which produce the original TDE \citep[an argument first put forward in][]{Linial2023}. 

We can however rule out this IMBH scenario based entirely on QPE observations, first by noting that $t_{\rm QPE}\sim P_{\rm QPE}^0$, in stark contrast with observations. Further,  by considering how much mass would be accreted onto the IMBH in each disk crossing. The Bondi accretion rate is given by 
\begin{equation}
\dot M_B \approx 4\pi R_{\rm bondi}^2 \rho_{\rm disk} |\vec v_{\rm rel}| \approx {4 \pi G^2 m_\bullet^2 \rho_{\rm disk} \over |\vec v_{\rm rel}|^3}, 
\end{equation}
meaning that in one orbital crossing $t_{\rm cross} \approx h/(v_K\cos(\psi))$, an amount of mass 
\begin{equation}
\Delta M_{\rm acc, \bullet} \approx t_{\rm cross} \dot M_B \approx  {\pi \Sigma_{\rm disk} R_{\rm QPE}^2} \left({m_\bullet \over M_\bullet}\right)^2 {[f(\psi)]^{3/2} \over \cos\psi} \approx M_{\rm disk} \left({m_\bullet \over M_\bullet}\right)^2 {[f(\psi)]^{3/2} \over \cos\psi}, 
\end{equation}
The high companion masses required to explain observed energies $m_\bullet \sim 10^4-10^5 M_\odot$ would then quickly drain the disk itself\footnote{We see that this problem is not resolved by fine tuning the IMBH orbit, as the lower mass for the IMBH needed in this limit is compensated by both the longer time it spends in the disk and the higher Bondi accretion rate for lower relative velocities.}. 

\subsection{ Collisions between stripped stellar debris and the disk }
The previous two sections show that neither a stellar surface or an IMBH can act as a colliding object and reproduce QPE-TDE observations.  As will be discussed in more detail below, the repeated crossings of the stellar object through the disk will lead to mass being stripped off the stellar EMRI, owing to the ram pressure of the collision. This will naturally act to increase the effective ``size'' of the colliding object, leading to a growing cross section available to future collisions. We take into account this effect in the following section. 

In the numerical simulations of \cite{Yao24} it was indeed shown that a stream of stellar debris is stripped off the star (of course some of this stream is also shocked disk material), which then increases the surface area for the next collision. \cite{Yao24} demonstrate with simplified analytical arguments, and full numerical simulations, that the length of this debris stream is given by
\begin{equation}
    {\Delta z \over R^{\rm emri}_\star} \sim \left({R_{\rm QPE} \over r_T^{\rm emri}}\right)^{3/2},
\end{equation}
while the dispersion in the debris in the plane of the disk is relatively limited 
\begin{equation}
    \Delta x \sim \Delta y \sim R_\star^{\rm emri},
\end{equation}
where we have introduced the debris stream coorindates in the disk plane $(x,y)$ and perpendicular to the disk $(z)$. This means that the cross sectional area hitting the disk at any one moment remains $A_{\rm hit} \sim \pi (R_\star^{\rm emri})^2$, but that the total surface area of stellar-debris-stream disk collisions are enhanced by a factor\footnote{Now that the debris stream is larger than the disk thickness it is likely that the collision will no longer produce symmetric ejecta, and there may be only one observable flare per orbit. If true, then the substitution $P_{\rm QPE}\to 2P_{\rm QPE}$ in the following expressions should be made.}
\begin{equation}
    {A_{\rm debris} \over \pi (R_\star^{\rm emri})^2} \sim {\Delta x \Delta z \over (R^{\rm emri}_\star)^2} \sim \left({R_{\rm QPE} \over r_T^{\rm emri}}\right)^{3/2} ,
\end{equation}
where 
\begin{equation}
    r_T^{\rm emri} \approx R_\star^{\rm emri} \left({M_\bullet \over M_\star^{\rm emri}}\right)^{1/3} , 
\end{equation}
is the tidal radius of the object on the EMRI. This enhancement can be thought of as a series of collisions with a surface $\sim \pi (R_\star^{\rm emri})^2$ which last a time $\Delta t \sim \Delta z/v_K$. Turning this area into an effective collisional radius (this neglects order unity constants which will not modify the scaling laws)
\begin{equation}
    A_{\rm debris} \approx \pi R_{\rm QPE}^{3/2} \left({M_\star^{\rm emri} R_\star^{\rm emri}\over M_\bullet}\right)^{1/2} \approx P_{\rm QPE} (GM_\star^{\rm emri} R_\star^{\rm emri})^{1/2},  
\end{equation}
or explicitly 
\begin{equation}
    {A_{\rm debris} \over \pi (R_\star^{\rm emri})^2} \approx 3  \left({P_{\rm QPE} \over 4 \, {\rm hours}}\right) \left({R_\star^{\rm emri} \over R_{\odot}}\right)^{-3/2}  \left({M_\star^{\rm emri} \over M_\odot}\right)^{1/2}  .
\end{equation}
We see that this is indeed a significant enhancement in the area, and that this enhancement grows with orbital period. This orbital period dependence will offset some of the energy drop inherent in the lower disk mass at these larger radii. 

This enhanced area increases the total kinetic energy in the collisions, to 
\begin{equation}
    E_{\rm ej} \approx 1.4 \times 10^{48} \, {\rm erg} \,  {f_d\over \beta^{1/2}} \left({M_\star^{\rm tde}\over M_\odot}\right)^{5/6} \left({R_\star^{\rm tde} \over R_{\odot}}\right)^{1/2} \left({R_\star^{\rm emri} \over R_{\odot}}\right)^{1/2} \left({M_\star^{\rm emri} \over M_\odot}\right)^{1/2}  \left({4 \, {\rm hours} \over P_{\rm QPE}}\right)^{4/3}. 
\end{equation}

The debris stream, in addition to creating an enhanced area for the collisions, also creates a spread in arrival times for the returning debris, approximately described by \citep{Yao24}
\begin{equation}
    \Delta t\sim {\Delta z\over v_K} \to {\Delta t \over P_{\rm QPE}} \approx 6\sqrt{2} \left({R_{\rm QPE} \over r_T^{\rm emri}}\right)^{1/2} \left({M_\star^{\rm emri} \over M_\bullet}\right)^{1/3} ,
\end{equation}
or explicitly 
\begin{equation}
    {\Delta t} \approx 0.5\, {\rm hours}\,  \left({P_{\rm QPE} \over 4\, {\rm hours}}\right)^{4/3} \left({M_\star^{\rm emri} \over M_\odot}\right)^{1/2}\left({R_\star^{\rm emri} \over R_\odot}\right)^{-1/2}\left({M_\bullet \over 10^6M_\odot}\right)^{-1/3} ,
\end{equation}
which, if it is this which sets the flare duration of the QPE, is more in keeping with the observed $t_{\rm QPE}-P_{\rm QPE}$ observations. This would require that it is this timescale, not the diffusion time for the photons to escape, which sets the duration of the event. This can be checked by explicit computation. 

There are two possible behaviors for a collision between a stream of stellar debris and an accretion flow, depending on  whether the surface density of the stellar debris stream $\Sigma_{\rm debris}$ is greater or lesser than the surface density of the disk $\Sigma_{\rm disk}$. This is difficult to ascertain on purely analytical grounds, but information from numerical simulations will help here. 

\cite{Yao24} argue that their simulations are well described by  
\begin{equation}\label{densrat}
    {\Sigma_{\rm debris} \over \Sigma_{\rm disk}} \sim 50 \left({M_\star^{\rm emri} \over M_\odot}\right) \left({M_\bullet \over 10^6 M_\odot }\right)^{2/3} \left({4 \, {\rm hours} \over P_{\rm QPE}}\right)^{-2} ,
\end{equation}
suggesting that while short period QPEs are in the $\Sigma_{\rm debris} \gg \Sigma_{\rm disk}$ limit, long period QPEs may well not be\footnote{The reader may be surprised that that this debris stream density can be higher than the disk density, while it is simultaneously true that adding this debris to the disk will not increase the disk density by any meaningful fraction. The reason this is true is because after one orbit (when the material returns to the disk) the debris is contained within a much smaller area than the area it will take up when it is spread out into the disk. This fraction can be calculated, and is equal to   $A_{\rm debris}/\pi R_{\rm QPE}^2 \approx 7 \times 10^{-5} \, (P_{\rm QPE}/4{\rm hours})^{-1/3} (R_\star^{\rm emri}/R_\odot)^{1/2} (M_\star^{\rm emri}/M_\odot)^{1/2} (M_\bullet / 10^6 M_\odot)^{-2/3}$. Once this debris  is spread out across the disk its contribution to the density drops by a factor $\sim 10^4$, consistent with the argument in previous sections.}.

In the limit where $\Sigma_{\rm debris} \gg \Sigma_{\rm disk}$ the collision proceeds much as discussed in the ``hard sphere'' limit (as above), only with an enhanced total area (now set by $\sim A_{\rm debris}$). Each shell of shocked gas caused by a collision between an element of the debris stream will radiate after a diffusion time equal to that derived earlier in the hard sphere limit (as the two collisions are mediated through the same colliding surface area)
\begin{equation}
    t_{\rm diff} \approx 1.1\, {\rm hours }\,  {f_d^{1/2} \over \beta^{1/4}} \left({M_\star^{\rm tde}\over M_\odot}\right)^{5/12} \left({R_\star^{\rm tde} \over R_{\odot}}\right)^{1/4}   \left({R_\star^{\rm emri} \over R_{\odot}}\right)  \left({4 \, {\rm hours} \over P_{\rm QPE}}\right)^{2/3} \left({10^6 M_\odot \over M_\bullet}\right)^{1/2} .
\end{equation}
%In this limit, working through $A_{\rm debris}$ in place of $\pi (R_\star^{\rm emri})^2$ from the earlier sections, we find that the observable parameters scale as 
% \begin{equation}
%     t_{\rm diff, debris} \approx 1.9\, {\rm hours }\,  {f_d^{1/2} \over \beta^{1/4}} \left({M_\star^{\rm tde}\over M_\odot}\right)^{5/12} \left({R_\star^{\rm tde} \over R_{\odot}}\right)^{1/4}   \left({R_\star^{\rm emri} \over R_{\odot}}\right)^{1/4}  \left({M_\star^{\rm emri} \over M_\odot}\right)^{1/4} \left({4 \, {\rm hours} \over P_{\rm QPE}}\right)^{1/6} \left({10^6 M_\odot \over M_\bullet}\right)^{1/2} ,
% \end{equation}
meaning that for the shortest period QPEs, the diffusion time would set the flare duration, while the stream length would set the flare duration for the majority of all  QPEs. The efficiency of turning the kinetic energy of the collision into radiation is set by the volume expansion factor 
\begin{equation}
    \lambda = \left({3V_0 \over 4\pi R_{\rm ej}^3}\right)^{1/3} = \left({3 (R_\star^{\rm emri})^2 h\over 28 v_K^3 t_{\rm diff}^3}\right)^{1/3},
\end{equation}
where we again use the fact that the initial collision is mediated through a surface area $A_{\rm hit} \sim \pi (R_\star^{\rm emri})^2$.  This is again equal to the efficiency of the initial hard sphere calculation, and so 
\begin{equation}
E_{\rm rad, debris} \approx 1.1 \times 10^{46} \, {\rm erg} \, {f_d^{1/2}\over \beta^{1/4}} \left({M_\star^{\rm tde}\over M_\odot}\right)^{5/12} \left({R_\star^{\rm tde} \over R_{\odot}}\right)^{1/4}\, \left(10{h\over r}\right)^{1/3}  
\left({R_\star^{\rm emri} \over R_{\odot}}\right)^{1/6}  \left({M_\star^{\rm emri} \over M_\odot}\right)^{1/2} \left({4 \, {\rm hours} \over P_{\rm QPE}}\right)^{1/9} \left({ M_\bullet \over 10^6 M_\odot}\right)^{-1/8} . 
\end{equation}
This flare would then have an average luminosity $L_{\rm flare} \approx \lambda E_{\rm ej}/\Delta t$, or 
% \begin{equation}
    % L_{\rm flare, debris} \approx 7.5 \times 10^{43} \, {\rm erg/s} {f_d\over \beta^{1/2}} \left({M_\star^{\rm tde}\over M_\odot}\right)^{5/6} \left({R_\star^{\rm tde} \over R_{\odot}}\right)^{1/2} 
    % \left({\lambda \over 0.1}\right) \left({R_\star^{\rm emri} \over R_{\odot}}\right) \left({M_\bullet\over 10^6 M_{\odot}}\right)^{1/3} \left({4 \, {\rm hours} \over P_{\rm QPE}}\right)^{8/3}. 
% \end{equation}
\begin{equation}
L_{\rm flare, debris} \approx 6.1 \times 10^{42} \, {\rm erg\, s}^{-1} \,{f_d^{1/2}\over \beta^{1/4}} \left({M_\star^{\rm tde}\over M_\odot}\right)^{5/12} \left({R_\star^{\rm tde} \over R_{\odot}}\right)^{1/4}\, \left(10{h\over r}\right)^{1/3}   \left({R_\star^{\rm emri} \over R_{\odot}}\right)^{2/3}   \left({4 \, {\rm hours} \over P_{\rm QPE}}\right)^{13/9} \left({ M_\bullet \over 10^6 M_\odot}\right)^{5/24} .
\end{equation}
This average luminosity drops rapidly with period, a scaling not seen in the data. 

% and 
% \begin{equation}
% k T_{\rm BB, debris} \approx 6.5 \, {\rm eV} \, {\beta^{1/8} \over f_d^{1/4}} \, \left(10{h\over r}\right)^{1/12} \left({P_{\rm QPE} \over 4\, {\rm hours}}\right)^{11/36} \, \left({M_\star^{\rm tde}\over M_\odot}\right)^{-5/24} \left({R_\star^{\rm tde} \over R_{\odot}}\right)^{-1/8} \left({R_\star^{\rm emri} \over R_{\odot}}\right)^{-1/12}  \left({M_\star^{\rm emri} \over M_\odot}\right)^{-1/12} \left({ M_\bullet \over 10^6 M_\odot }\right)^{-1/8} . 
% \end{equation}

As the collision occurs through the same surface area as in the hard sphere limit $(A_{\rm hit} \sim \pi (R_\star^{\rm emri})^2)$, the energy density of the shocked gas after one photon diffusion time is unchanged from the hard sphere calculation, meaning that the blackbody temperature of the radiation is unchanged from above.  Note that again the properties of the radiation field (i.e., the effects of photon starvation and inverse Comptonization) are unchanged in this scenario. While the increased area of the stripped debris streams does increase the energy and time period at longer period collisions, we still find a negative scaling with QPE period for both the flare luminosity and flare energy. 

In the opposite limit $\Sigma_{\rm disk}\gg \Sigma_{\rm debris}$, there will in principle be two emission components \citep{Yao24}. The first will be the star punching through the disk (the hard sphere calculation discussed above), and the second being the shock heating of the stellar debris stream by its collision with the disk (while this debris stream has minimal impact on the properties of the disk, it will itself be heated to high temperatures). Given that, by definition, $A_{\rm debris} \Sigma_{\rm debris} \ll A_{\rm debris} \Sigma_{\rm disk}$ in this limit, and that the disk material is orbiting with the same bulk velocity $(v_{\rm disk} \simeq v_K \simeq v_{\rm EMRI} )$ as the EMRI, it seems that the energy injected into the orbiting debris stream by this collision $E_{\rm kin, debris} \sim A_{\rm debris} \Sigma_{\rm debris} v_{\rm disk}^2 \ll A_{\rm debris} \Sigma_{\rm disk} v_{\rm EMRI}^2$ is significantly lower than that calculated in the above calculation. As the above calculation does not appear to generate enough energy to explain the longest period QPE observations, we take it as an optimistic upper bound for either debris stream-disk collision scenario. 

We do not consider the case $\Sigma_{\rm disk} \sim \Sigma_{\rm debris}$ in this paper, which will clearly be more complex than either limit. It will likely require numerical radiative transfer simulations \citep[\`a la][]{Vurm24} to deal with this limit properly.

% If it is this longer timescale that sets the flare duration, then perhaps a reasonable fraction of the kinetic energy of the collision can be liberated over $\Delta t$. It becomes harder to estimate the exact radiated energy in this limit, but defining an efficiency $\lambda$ by 
% \begin{equation}
%     \lambda \equiv {E_{\rm rad, \Delta t} \over E_{\rm kin}} ,
% \end{equation}
% we have 
% \begin{align}
%     E_{\rm rad, \Delta t} &\approx \lambda A_{\rm debris} \Sigma_{\rm disk} v_{K}^2 \\
%     &\approx 1.3 \times 10^{47} \, {\rm erg} \,  {f_d\over \beta^{1/2}} \left({M_\star^{\rm tde}\over M_\odot}\right)^{5/6} \left({R_\star^{\rm tde} \over R_{\odot}}\right)^{1/2} 
%     \left({\lambda \over 0.1}\right) \left({R_\star^{\rm emri} \over R_{\odot}}\right)^{1/2} \left({M_\star^{\rm emri} \over M_{\odot}}\right)^{1/2} \left({4 \, {\rm hours} \over P_{\rm QPE}}\right)^{4/3}. 
% \end{align}

\subsection{A ``puffed up'' stellar EMRI }
An upper limit on the spherical size of the colliding object will be set by the volume of the EMRI's Hills sphere\footnote{I am grateful to Brian Metzger for suggesting this calculation.}, which roughly sets the volume over which stellar material will remain gravitationally bound to the star and not the central supermassive black hole. Numerical simulations of the star-disk interaction seem to suggest that the stellar debris relatively quickly fill its Hills sphere \citep{Yao24}. 

Approximately, the size of the Hills sphere is given by 
\begin{equation}
    R_H \approx R_{\rm QPE} \left({M_\star^{\rm emri} \over M_\bullet}\right)^{1/3} \approx \left({P_{\rm QPE} \over \pi}\right)^{2/3} \left(GM_\star^{\rm emri}\right)^{1/3} ,
\end{equation}
which can be verified in the standard manner by computing the distance from the EMRI at which the sum of the gravitational forces towards the EMRI and black hole cancel the centripetal acceleration of an orbit with angular frequency $\Omega^2 \approx GM_\bullet / R_{\rm QPE}^3$. In reality this expression will be modified by  order unity prefactors (which of course cannot change the scaling laws derived here).  

Evaluating the above radial size, we find 
\begin{equation}
    R_H \approx 2 R_\odot \left({P_{\rm QPE} \over 4 \, {\rm hours}}\right)^{2/3} \left({M_\star^{\rm emri}\over M_\odot}\right)^{1/3}  ,
\end{equation}
where we see that this Hills sphere is typically $\sim 1-10$ times larger than the radius of the star itself, for $P_{\rm QPE} \sim 2-72$ hours (the observed period range of QPE sources; Figure \ref{fig:data}). Feeding this new scaling through to the results derived above we find that the observable parameters scale as 
\begin{equation}
    t_{\rm diff, Hills} \approx 2.1\, {\rm hours }\,  {f_d^{1/2} \over \beta^{1/4}} \left({M_\star^{\rm tde}\over M_\odot}\right)^{5/12} \left({R_\star^{\rm tde} \over R_{\odot}}\right)^{1/4}   \left({M_\star^{\rm emri} \over M_{\odot}}\right)^{1/3}  \left({10^6 M_\odot \over M_\bullet}\right)^{1/2} ,
\end{equation}
\begin{equation}
E_{\rm rad, Hills} \approx 1.1 \times 10^{46} \, {\rm erg} \, {f_d^{1/2}\over \beta^{1/4}} \left({M_\star^{\rm tde}\over M_\odot}\right)^{5/12} \left({R_\star^{\rm tde} \over R_{\odot}}\right)^{1/4}\, \left(10{h\over r}\right)^{1/3}  
\left({M_\star^{\rm emri} \over M_{\odot}}\right)^{5/9}  \left({ M_\bullet \over 10^6 M_\odot}\right)^{1/8} , 
\end{equation}
\begin{equation}
L_{\rm flare, Hills} \approx 1.5 \times 10^{42} \, {\rm erg\, s}^{-1} \,\left(10{h\over r}\right)^{1/3}  \left({M_\star^{\rm emri} \over M_{\odot}}\right)^{2/9}  \left({ M_\bullet \over 10^6 M_\odot}\right)^{5/8} , 
\end{equation}
and 
\begin{equation}
k T_{\rm BB, Hills} \approx 6.3 \, {\rm eV} \, {\beta^{1/8} \over f_d^{1/4}} \,  \left({M_\star^{\rm tde}\over M_\odot}\right)^{-5/24} \left({R_\star^{\rm tde} \over R_{\odot}}\right)^{-1/8} \left(10{h\over r}\right)^{1/12} \left({P_{\rm QPE} \over 4\, {\rm hours}}\right)^{1/4} \,\left({M_\star^{\rm emri} \over M_{\odot}}\right)^{-1/9}  \left({ M_\bullet \over 10^6 M_\odot }\right)^{-1/8} . 
\end{equation}
We note some of the similarities in these scalings with the case of an intermediate mass black hole on the EMRI, in particular we now have that neither the duration, luminosity  or the energy of the QPE flares are predicted to scale with the QPE period. Just as above in the IMBH case, the properties of the radiation field (i.e., the effects of photon starvation and inverse Comptonization) are unchanged in this scenario from that derived for a ``hard sphere'' stellar EMRI, as they depend only on the properties of the disrupted star.

Of course, for a long period orbit, this Hills sphere can be significantly larger than the star itself, and there will be a spread in arrival times of the orbiting debris of 
\begin{equation}
    {\Delta t\over P} \approx {R_H \over R_{\rm QPE}} \approx \left({M_\star^{\rm emri}\over M_\bullet}\right)^{1/3} \approx 0.01 \left({M_\star^{\rm emri}\over M_\odot}\right)^{1/3} \left({10^6 M_\odot\over M_\bullet}\right)^{1/3} .
\end{equation}
While this linear scaling is in good accord with the data,  the amplitude is much smaller than the photon diffusion timescale, and so it is this photon diffusion timescale which sets the flare duration. We therefore believe that this Hills sphere model is ruled out by the $t_{\rm QPE}-P_{\rm QPE}$ relationship. 
%If we again assume that over $\Delta t$ some fraction $\lambda$ of the kinetic energy of this Hills-sphere collision is converted into X-ray emission, then 
% \begin{equation}
%     E_{\rm rad, \Delta t, Hills} \approx 1.8\times 10^{47} \, {\rm erg} \,  {f_d\over \beta^{1/2}} \left({M_\star^{\rm tde}\over M_\odot}\right)^{5/6} \left({R_\star^{\rm tde} \over R_{\odot}}\right)^{1/2}   \left({\lambda \over 0.1}\right) \left({M_\star^{\rm emri} \over M_{\odot}}\right)^{2/3} \left({4 \, {\rm hours} \over P_{\rm QPE}}\right) . 
% \end{equation}
% Again however, this radiated energy drops with orbital period, leading to a rapidly dropping average luminosity
% \begin{equation}
%     L_{\rm QPE, \Delta t, Hills} \approx 1.3 \times 10^{44} {\rm erg/s}\,   {f_d\over \beta^{1/2}} \left({M_\star^{\rm tde}\over M_\odot}\right)^{5/6} \left({R_\star^{\rm tde} \over R_{\odot}}\right)^{1/2}   \left({\lambda \over 0.1}\right) \left({M_\star^{\rm emri} \over M_{\odot}}\right)^{1/3} \left({10^6 M_\odot\over M_\bullet}\right)^{1/3}  \left({4 \, {\rm hours} \over P_{\rm QPE}}\right)^{2}. 
% \end{equation}

\subsection{A hybrid Hills sphere--debris stream model }
The previous two subsections discussed flare properties in the limit in which the stellar EMRI has been collisionally unbound in some sense, leading to an enhanced collisional cross section with the disk. We first considered the impact of an extended debris stream, and then the modifications resulting from the EMRI being puffed up to fill its Hills sphere. Both of these modifications can be taken as inspired by the numerical simulations of \cite{Yao24}, but we have shown in this paper that they would be required for the collisional paradigm to work, independent of these simulations, on purely energetic grounds.

There is, however, no reason that collisions between an accretion disk and an EMRI with long periods cannot by described by a hybrid version of the two models, where the EMRI has been puffed up to fill its Hills sphere and is also being trailed by a debris stream.  This in some sense is a maximally optimistic model in terms of generating enough energy in the collisions to explain the observations of long period QPEs. 

In this model the length of this debris stream is given by \citep{Yao24}
\begin{equation}
    {\Delta z \over R_\star} \sim \left({R_{\rm QPE} \over r_T^{\rm emri}}\right)^{3/2},
\end{equation}
while the dispersion in the debris in the plane of the disk is set by the Hills sphere 
\begin{equation}
    \Delta x \sim \Delta y \sim R_H .
\end{equation}
This implies an enhanced collision surface area of 
\begin{equation}
    {A_{\rm debris} \over \pi (R_\star^{\rm emri})^2} \sim {\Delta x \Delta z \over (R^{\rm emri}_\star)^2} \sim \left({R_{\rm QPE} \over r_T^{\rm emri}}\right)^{5/2} .
\end{equation}
We therefore have 
\begin{equation}
    A_{\rm debris} \approx \pi R_{\rm QPE}^{2} \left({R_{\rm QPE} \over R_\star^{\rm emri}}\right)^{1/2} \left({M_\star^{\rm emri} \over M_\bullet}\right)^{5/6} \approx \pi^{-2/3} P_{\rm QPE}^{5/3} \left(GM_\star^{\rm emri}\over  (R_\star^{\rm emri})^{3/5}\right)^{5/6},  
\end{equation}
or explicitly 
\begin{equation}
    {A_{\rm debris} \over \pi (R_\star^{\rm emri})^2} \approx 5.8  \left({P_{\rm QPE} \over 4 \, {\rm hours}}\right)^{5/3} \left({R_\star^{\rm emri} \over R_{\odot}}\right)^{-1/2}  \left({M_\star^{\rm emri} \over M_\odot}\right)^{5/6}  .
\end{equation}
Folding  this area through gives a kinetic energy in the collision debris of 
\begin{equation}
    E_{\rm ej} \approx 2.7 \times 10^{48} \, {\rm erg} \,  {f_d\over \beta^{1/2}} \left({M_\star^{\rm tde}\over M_\odot}\right)^{5/6} \left({R_\star^{\rm tde} \over R_{\odot}}\right)^{1/2} \left({M_\star^{\rm emri} \over M_{\odot}}\right)^{5/6}\left({R_\star^{\rm emri} \over R_{\odot}}\right)^{3/2} \left({4 \, {\rm hours} \over P_{\rm QPE}}\right)^{2/3}. 
\end{equation}
As the collision is mediated through a cross section of area $\pi R_H^2$, we can reuse the above Hills sphere results for the diffusion time and radiation temperature. The efficiency of turning kinetic energy into radiation is then given by 
\begin{equation}
    \lambda = \left({3V_0 \over 4\pi R_{\rm ej}^3}\right)^{1/3} = \left({3 (R_H)^2 h\over 28 v_K^3 t_{\rm diff, Hills}^3}\right)^{1/3},
\end{equation}
which equals 
\begin{equation}
    \lambda \approx 6.2 \times 10^{-3} \, {f_d^{-1/2} \over \beta^{-1/4}} \left({M_\star^{\rm tde}\over M_\odot}\right)^{-5/12} \left({R_\star^{\rm tde} \over R_{\odot}}\right)^{-1/4}   
    \left(10{h\over r}\right)^{1/3}  \left({M_\star^{\rm emri} \over M_{\odot}}\right)^{-1/3} \left({P_{\rm QPE} \over 4\, {\rm hours}}\right) .
\end{equation}
The radiated energy is then 
\begin{equation}
    E_{\rm rad} \approx 1.6 \times 10^{46} \, {\rm erg} \,  {f_d^{1/2}\over \beta^{1/4}} \left({M_\star^{\rm tde}\over M_\odot}\right)^{5/12} \left({R_\star^{\rm tde} \over R_{\odot}}\right)^{1/4} 
    \left(10{h\over r}\right)^{1/3} \left({M_\star^{\rm emri} \over M_{\odot}}\right)^{1/2}\left({R_\star^{\rm emri} \over R_{\odot}}\right)^{3/2} \left({P_{\rm QPE} \over 4 \, {\rm hours} }\right)^{1/3}. 
\end{equation}
The debris stream again sets the length of duration of this flare 
\begin{equation}
    {\Delta t} \approx 0.5\, {\rm hours}\,  \left({P_{\rm QPE} \over 4\, {\rm hours}}\right)^{4/3} \left({M_\star^{\rm emri} \over M_\odot}\right)^{1/2}\left({R_\star^{\rm emri} \over R_\odot}\right)^{-1/2}\left({M_\bullet \over 10^6M_\odot}\right)^{-1/3} ,
\end{equation}
meaning an average flare luminosity 
\begin{equation}
    L_{\rm flare} \approx 9.4 \times 10^{42} \, {\rm erg/s}\,  {f_d^{1/2}\over \beta^{1/4}} \left({M_\star^{\rm tde}\over M_\odot}\right)^{5/12} \left({R_\star^{\rm tde} \over R_{\odot}}\right)^{1/4} 
    \left(10{h\over r}\right)^{1/3} \left({M_\bullet \over 10^6 M_{\odot}}\right)^{1/3}\left({R_\star^{\rm emri} \over R_{\odot}}\right)^{2} \left({P_{\rm QPE} \over 4 \, {\rm hours} }\right)^{-1}. 
\end{equation}
These results are best in keeping with the observations. There is an energy scale which grows with orbital period, and broadly the right amplitude at shorter periods (Figure \ref{fig:data}). However, we remind the reader that this is a highly elongated structure (in effect an absolute upper bound), and we have implicitly assumed that this structure has a surface density $\Sigma_{\rm debris}\gg \Sigma_{\rm disk}$, so that it effectively punches through the disk. Given the highly elongated structure required, it is not immediately clear that such a density scale can be achieved, except for collisions between a disk and a particularly large mass star (e.g., equation \ref{densrat}, and discussion in \citealt{Yao24}). 

\subsection{ Subsequent time evolution }
The above analysis evaluates the properties of the emission from the collisionally shocked gas at the time at which the disk first expands to a large enough radial extent that it intercepts the EMRI orbital radius. This of course does not necessarily correspond to the first time at which emission would be first detected from such a system. %Indeed, the above scaling relationships typically imply that emission will not be detectable at first for short period QPEs (the inferred observable temperatures of the emission are too soft). 

At  times beyond this first collision, there are three interacting processes which will determine the evolution of the system.  The first is that the original TDE disk will of course continue to accrete through its inner edge, which will generally act to lower the density of the material which the object on an EMRI is colliding with. However, counteracting this effect is the fact that collisions between the disk and EMRI will strip mass from the object (assuming that it is not a black hole), which will generally increase the local density of the accretion flow. Thirdly, presumably some of the ejected material from the EMRI-disk collision will be unbounded and escape, which will gradually deplete the mass content of the disk. The rates of these three mass loss terms are calculated below. The mass loss from the disk due to the unbinding of the shocked debris is 
\begin{equation}
\dot M_{\rm eject} = - f_{\rm u} {M_{\rm ej} \over P_{\rm QPE}} ,
\end{equation}
or explicitly 
\begin{equation}
\dot M_{\rm eject} \approx - 0.05 M_\odot \, {\rm yr}^{-1}  {f_uf_d \over \beta^{1/2}} \left({M_\star^{\rm tde}\over M_\odot}\right)^{5/6}  \left({R_\star^{\rm tde} \over R_{\odot}}\right)^{1/2} \left({R_{\rm col} \over R_{\odot}}\right)^2\left({4 \, {\rm hours} \over P_{\rm QPE}}\right)^{8/3} \left({10^6 M_\odot \over M_\bullet}\right)^{2/3} \left({t \over \Delta t_{\rm QPE}}\right)^{-n} ,
\end{equation}
where $f_{\rm u}$ is the fraction of material unbound in each collision, and the time dependence is inherited from the density evolution (assuming an accretion dominated flow).  It is unclear precisely what value $f_u$ should take, but it is almost certainly non-zero as the shocked debris possesses a characteristic spread in its specific internal energy $\sim v_K^2$ comparable to the escape speed. If $f_u \ll 1$ this  effect is unimportant (relative to accretion and ablation) except for low black hole masses and QPE periods, although if $f_u \sim 1$ this effect can be important, particularly if the effective collisional radius $R_{\rm col} \sim \sqrt{A_{\rm col}/\pi}$ is super solar\footnote{The mass ejection timescale can be computed for the various collision scenarios discussed above upon substitution of the Bondi radius $R_{B}$, effective debris radius $R_{\rm debris}=\sqrt{A_{\rm debris}/\pi}$  or Hills radius $R_{H}$ (and their various parameter dependencies) into the above expression in place of $R_{\rm col}$.}.   As $f_u$ is poorly constrained {\it apriori}, we shall neglect mass unbinding in the following discussions.

The accretion rate through the inner edge of the disk will be approximately given by 
\begin{equation}
\dot M_{\rm acc} \simeq - (n - 1) {M_{\rm disk}^{\rm QPE} \over \Delta t_{\rm QPE}} \left({t \over \Delta t_{\rm QPE}}\right)^{-n}, \quad t > \Delta t_{\rm QPE} , 
\end{equation}
where $\Delta t_{\rm QPE}$ is the time taken for the disk to first spread to intercept the EMRI's orbital radius after the first disruption. The normalisation here simply encodes the fact that at times $t > \Delta t_{\rm QPE}$ all of the material that is left at this point $(M_{\rm disk}^{\rm QPE})$ would eventually be accreted in the absence of an intercepting EMRI (which doubtless complicates matters). The index $n$ of this self similar solution depends on precise assumptions about the turbulent stress in the disk, but is typically $n \simeq 1.2$ \citep{Cannizzo90}. This evaluates to 
\begin{equation}
\dot M_{\rm acc} \approx - 0.1 M_\odot \, {\rm yr}^{-1} \, \left({1\, {\rm year} \over \Delta t_{\rm QPE}}\right)  \, {f_d \over \beta^{1/2}}  \left({M_\star^{\rm tde}\over M_\odot}\right)^{5/6} \left({R_\star^{\rm tde} \over R_{\odot}}\right)^{1/2} \left({4\, {\rm hours} \over P_{\rm QPE}}\right)^{1/3} \left({t \over \Delta t_{\rm QPE}}\right)^{-n} .
\end{equation}
Note that this drops quite slowly with orbital period $\sim P_{\rm QPE}^{-1/3}$.

Finally, as the orbiting object is crashing through a disk, and therefore experiencing a ram pressure 
\begin{equation}
p_{\rm ram} \simeq \rho_{\rm disk} v_K^2  ,
\end{equation}
mass will be stripped upon each collision.  The mass stripped in each collision $\Delta M_{\rm emri}^{(1)}$ is expected to be equal to \citep{Linial2023}
\begin{equation}
\Delta M^{(1)}_{\rm emri} \approx \eta \left({p_{\rm ram} \over p^{\rm emri}_\star}\right) M_\star^{\rm emri} , 
\end{equation}
were $p_\star^{\rm emri} \approx G(M_\star^{\rm emri})^2 / 4\pi (R_\star^{\rm emri})^4$ is the average internal pressure of the EMRI, and the prefactor $\eta \approx 10^{-3}$ is inferred from the simulations of \cite{Liu15} \citep[note however that a higher mass stripping rate was found in][]{Yao24}. This equals 
\begin{multline}
\Delta M^{(1)}_{\rm emri} \approx 2.3\times10^{-5} M_\odot  \left({\eta \over 10^{-3}}\right) {f_d \over \beta^{1/2}}  \left({M_\star^{\rm tde} \over M_\odot}\right)^{5/6}   \left({R_\star^{\rm tde} \over R_\odot}\right)^{1/2}  \\   \left({M_\star^{\rm emri} \over M_\odot}\right)^{-1}   \left({R_\star^{\rm emri} \over R_\odot}\right)^{4} \left(10{h\over r}\right)^{-1} \left({M_\bullet \over 10^6 M_\odot}\right)^{-1/3} \left({P_{\rm QPE} \over 4\, {\rm hours}} \right)^{-3} \left({t \over \Delta t_{\rm QPE}}\right)^{-n} ,
\end{multline}
where the time dependence is inherited from the evolving disk surface density through the ram pressure (Appendix \ref{app:evolve} for a proof of this time dependence). This mass stripping therefore occurs at a rate 
\begin{equation}
\dot M_{\rm abl} \approx \eta \left({p_{\rm ram} \over p^{\rm emri}_\star}\right) {M_\star^{\rm emri} \over P_{\rm QPE}}, 
\end{equation}
which evaluates to 
\begin{multline}
\dot M_{\rm abl}  \approx +  0.05 M_\odot  \, {\rm yr}^{-1} \left({\eta \over 10^{-3}}\right) {f_d \over \beta^{1/2}}  \left({M_\star^{\rm tde} \over M_\odot}\right)^{5/6}   \left({R_\star^{\rm tde} \over R_\odot}\right)^{1/2}   \\
\left({M_\star^{\rm emri} \over M_\odot}\right)^{-1}   \left({R_\star^{\rm emri} \over R_\odot}\right)^{4} \left(10{h\over r}\right)^{-1} \left({M_\bullet \over 10^6 M_\odot}\right)^{-1/3} \left({P_{\rm QPE} \over 4\, {\rm hours}} \right)^{-4} \left({t \over \Delta t_{\rm QPE}}\right)^{-n}  .
\end{multline}
We see that for short period EMRI orbits this can in principle exceed the disk mass loss via accretion, meaning that the disk mass could plateau, or even grow, in mass (from its small late time value) due to disk-EMRI interactions, rather than simply decay with time. We stress again, however, that adding this mass cannot ultimately raise the {\it maximum} disk density reached during the evolution by more than a factor $\sim 2$ from the value of the original TDE disk surface density at the point at which it first crossed the EMRI orbit (i.e., the density used in all of the calculations in the previous sections, and the relevant scale for computing the energetics of the collisions). This is irrespective of the ``adding rate'', as both the TDE disk and the disk formed from the addition of stripped EMRI material have upper bounds on their mass at the same $\sim$ solar, scale (i.e., they would both represent disks which formed from stars). To see this more clearly consider the surface density addition rate from ablation 
\begin{align}
    \dot \Sigma_{\rm add} &\approx {\dot M_{\rm abl} \over \pi R_{\rm QPE}^2} \\ 
    &\approx 5 \times 10^{-3} \, {\rm g/cm}^2{\rm /s} \left({\eta \over 10^{-3}}\right) {f_d \over \beta^{1/2}}  \left({M_\star^{\rm tde} \over M_\odot}\right)^{5/6}   \left({R_\star^{\rm tde} \over R_\odot}\right)^{1/2}    \left({M_\star^{\rm emri} \over M_\odot}\right)^{-1}   \left({R_\star^{\rm emri} \over R_\odot}\right)^{4} \left(10{h\over r}\right)^{-1} \left({M_\bullet \over 10^6 M_\odot}\right)^{-1} \left({P_{\rm QPE} \over 4\, {\rm hours}} \right)^{-16/3} ,
\end{align}
which is a small addition rate (contrast with equation \ref{densTDE} for the density scale of TDE disk when it first intercepted the EMRI orbit) which drops rapidly with orbital period $\sim P_{\rm QPE}^{-16/3}$. 
% which we can contrast with the surface density loss from accretion. At fixed radius the surface density of the disk drops with time according to $\Sigma_{\rm disk}(R_{\rm QPE}, t) = \Sigma_{\rm disk}(R_{\rm QPE}, \Delta t_{\rm QPE}) (t/\Delta t_{\rm QPE})^{-n}$ (see Appendix \ref{app:evolve} for a proof), meaning that 
% \begin{align}
%     \dot \Sigma_{\rm disk, acc} &= -n{\Sigma_{\rm disk}\over \Delta t_{\rm QPE}} \left({t\over \Delta t_{\rm QPE}}\right)^{-n-1} , \\
%     &\approx - 6\times10^{-2}\, {\rm g/cm}^2{\rm /s} \, \left({1\, {\rm year} \over \Delta t_{\rm QPE}}\right)  {f_d \over \beta^{1/2}} \left({M_\star^{\rm tde}\over M_\odot}\right)^{5/6}   \left({R_\star^{\rm tde} \over R_{\odot}}\right)^{1/2} \left({4 \, {\rm hours} \over P_{\rm QPE}}\right)^{5/3} \left({10^6 M_\odot \over M_\bullet}\right)^{2/3} \left({t\over \Delta t_{\rm QPE}}\right)^{-n-1} .
% \end{align}
The time scale for EMRI destruction $\tau_{\rm abl} = M_\star^{\rm emri}/\dot M_{\rm abl}$ bounds the lifetime of the system and is equal to 
\begin{equation}
    \tau_{\rm abl} \approx 10\, {\rm yr} \, \left({\eta \over 10^{-3}}\right)^{-1} \left({M_\star^{\rm emri} \over M_\odot}\right)^{2}  \left({R_\star^{\rm emri} \over R_\odot}\right)^{-4} \left({M_\star^{\rm tde} \over M_\odot}\right)^{-5/6} \left({R_\star^{\rm tde} \over R_\odot}\right)^{-1/2}    {\beta^{1/2} \over f_d } \left(10{h\over r}\right) \left({M_\bullet \over 10^6 M_\odot}\right)^{1/3} \left({P_{\rm QPE} \over 4\, {\rm hours}} \right)^{4} .
\end{equation}
We note for short period QPEs occurring in a disk formed from a TDE (i.e., GSN 069-like systems), that this can be an extremely short timescale. Let us consider an optimistic  limit in which the disk grows in density over a time $\Delta t = f_{\rm abl}\tau_{\rm abl}$ from mass added due to the the EMRI-disk interaction (we take the factor $f_{\rm abl}\approx 1/10$ as fiducial so that there is still an orbiting body left to collide with the disk), without any of this extra mass being lost into the black hole. Then the extra surface density of the disk will be
\begin{align}
    \Sigma_{\rm disk, EMRI} &\approx f_{\rm abl} \tau_{\rm abl} \dot \Sigma_{\rm add} , \\
    &\approx 1.6 \times 10^{5} \, {\rm g/cm^2}\, \left({f_{\rm abl}\over 0.1}\right) \left({M_\star^{\rm emri} \over M_\odot}\right)\left({P_{\rm QPE} \over 4\, {\rm hours}} \right)^{-4/3} \left({10^6 M_\odot \over M_\bullet}\right)^{-2/3} .
\end{align}
If we contrast this surface density with that of the disk originally formed from the TDE when it first intercepted the EMRI orbit
\begin{equation}
\Sigma^{\rm QPE}_{\rm disk, TDE} \approx 1.6\times10^6 \, {\rm g/ cm}^{2} \, {f_d \over \beta^{1/2}} \left({M_\star^{\rm tde}\over M_\odot}\right)^{5/6}   \left({R_\star^{\rm tde} \over R_{\odot}}\right)^{1/2} \left({P_{\rm QPE} \over 4\, {\rm hours}} \right)^{-5/3} \left({10^6 M_\odot \over M_\bullet}\right)^{-2/3}  ,
\end{equation}
we see that this remains insufficient to alleviate the general ``TDE disk densities are low for high periods'' problem faced by TDE disks and long period QPEs. At low orbital periods, where $\tau_{\rm abl}$ is rather short and the TDE disk is still early into its evolution, the subsequent time evolution will be very complex and interesting.

What may well happen, and in fact is extremely likely, is that this mass addition rate will prevent the evolution of the disk density simply dropping at some point time, and will thereafter modify the simple accretion driven evolution of the TDE-EMRI disk system. An example of this evolution from very low densities was considered in \cite{Linial24}.  For typical TDE, EMRI and QPE parameters these mass gain and loss terms are comparable at late times, and so the long term evolution of the system is likely to be complex and beyond the level of the simple analytical arguments made thus far. This is especially true as the mass loss to accretion and mass gain from the EMRI stripping are coupled, as the mass stripping rate is set by the ram pressure of the disk, which is proportional to the density. Eventually, as $\Sigma_{\rm disk, TDE} \propto t^{-n}$ (Appendix \ref{app:evolve}), we have $\dot \Sigma_{\rm add} \sim t^{-n}$ and $\dot \Sigma_{\rm disk, TDE} \sim t^{-n-1}$, meaning that in the very long time limit the rate of change of the local surface density will become dominated by mass stripping in collisions.   In the limit in which mass loss via accretion dominates (likely the relevant limit for longer periods, and intermediate evolutionary times, which is the relevant parameter space for the observed systems AT2019qiz and AT2022upj), time dependent solutions for each of the observables can be derived, which we do in Appendix \ref{app:evolve}. 

% In this section we have reverted to the ``hard sphere'' limit when calculating explicit timescales. We remind the reader that the mass ejection timescale can be computed upon substitution of the Bondi radius $R_{B}$, effective debris radius $R_{\rm debris}=\sqrt{A_{\rm debris}/\pi}$  or Hills radius $R_{H}$ (and their various parameter dependencies) into the above expressions in place of $R_{\rm col}$ (while of course remembering that a black hole will not be ablated). Generally speaking, there does not appear to be any obvious limit in which the various evolutionary timescales are not comparable, and therefore there does not appear to be a limit in which the subsequent evolution of the system is easy to compute. It appears that numerical simulations \`a la \cite{Yao24} will be required to evolve the system forward in time. 

\section{ Comparison to observations }\label{obs}
In this section we perform a brief comparison to observations of QPE sources. We do not perform a detailed comparison to all known QPEs, focusing instead on AT2019qiz and the broader population-level properties of QPE sources. 
\subsection{ AT2019qiz }
AT2019qiz is one of the longest period QPE source to date, and is also a source unambiguously known to also be a TDE \citep{Nicholl19}. This source is therefore the most constraining for QPE models in the high-period limit. In this sub-section we take the values $P_{\rm QPE} = 48$ hours, and $M_\bullet = 2 \times 10^6 M_\odot$ \citep{Nicholl24}. We shall also fix $h/r = 0.1$, although the scale height of the disk only ever enters through weak powers (and could in principle be calculated as we have constraints on the inner disk temperature, see Appendix \ref{app:height}). This scale hight does not really modify the results presented here at all.  

\subsubsection{ Collisions with stellar surfaces or an IMBH } 
We begin with the properties of the first two models, namely collisions with a hard stellar surface or (for completeness) also the Bondi radius of an IMBH. 

The peak luminosity of the QPE flares in AT2019qiz is observed to be $L_{\rm QPE} \simeq 1.8\times 10^{43}$ erg/s, and the duration of the flares was on average $t_{\rm QPE} \simeq 9$ hours \citep{Nicholl24}. The temperature of the emission was measured to be $k T_{\rm obs} \simeq 110$ eV. We have three observables and (naively) four unknowns, namely the radius of the object on an EMRI, the mass of the disrupted star, the radius of the disrupted star and a nuisance parameter relating to the properties of the disruption itself: $f_d/\beta^{1/2}$. At first it appears that this is an under-constrained problem, however it transpires  that all observable quantities in the problem only depend on the product 
\begin{equation}
Q \equiv {f_d \over \beta^{1/2}} \left({M_\star^{\rm tde}\over M_\odot}\right)^{5/6} \left({R_\star^{\rm tde} \over R_{\odot}}\right)^{1/2} , 
\end{equation}
and we therefore have an {\it over-constrained} problem, which is more interesting. 

The luminosity of the QPE depends only on the radius of the EMRI, and we can therefore measure it directly in this interpretation 
\begin{equation}
R_\star^{\rm emri} \approx 249 R_\odot, \quad m_\bullet \approx 3.9 \times 10^5 M_\odot,
\end{equation}
where we also compute the IMBH mass required for this to be the Bondi radius\footnote{We note that if the orbit of the IMBH was highly fine-tuned so that it was orbiting  in the same direction as, and just misaligned with, the disk, then this mass is reduced by at most a factor $m_\bullet \to m_\bullet(h/r)^{2}$. This is a very unnatural solution to the problem, and cannot reasonably also be invoked for AT2022upj, another QPE-TDE with a similar period and observed energy.}. The stellar radius seems implausible\footnote{Note that this is not unique to the TDE disk model developed here and other published models would require an even larger radius $R_\star^{\rm emri} \approx 2500 R_\odot$ which is bigger than the EMRI orbit (this larger collisional radius results from the much thinner disk implied in the AGN disk model at this large radius). } if we are to interpret this as an orbiting star (this radius represents roughly 20\% of the orbital radius which the object would have to be on to collide at $R_{\rm QPE}$, and would in fact be tidally disrupted itself at the orbital radius $R_{\rm QPE}$ for masses $M_{\star}^{\rm emri} \lesssim 13,000M_\odot$). Of course the IMBH mass is physically possible, if potentially unlikely (see earlier discussion about the growth rate of SMBHs if this many $m_\bullet \sim 10^5 M_\odot$ IMBHs are in existence around a typical SMBH). With the orbiting radius constrained, we can further constrain the mass and radius of the disrupted star from the timescale of the emission. Requiring that the timescale match the observed value provides the following  constraint on the product $Q$ 
\begin{equation}
Q^{1/2} \approx 0.25 \approx {f_d^{1/2}\over \beta^{1/4}} \left({M_\star^{\rm tde}\over M_\odot}\right)^{5/12} \left({R_\star^{\rm tde} \over R_{\odot}}\right)^{1/4} , 
\end{equation}
If we were to assume a simple mass-radius relationship of the form $R_\star^{\rm tde} = R_\odot (M_\star^{\rm tde} / M_\odot)^{4/5}$, this would imply 
\begin{equation}
M_\star^{\rm tde} \approx 0.1 \left({\beta^{15/37}\over f_d^{30/37}}\right) M_\odot \gtrsim 0.1 M_\odot, 
\end{equation}
where the second inequality follows as $\beta > 1$ and $f_d < 1$. This star would have a tidal radius about an AT2019qiz black hole of mass $M_\bullet = 2 \times 10^6 M_\odot$ of  
\begin{equation}
r_T^{\rm tde} \approx 10 \left({\beta^{7/37}\over f_d^{14/37}}\right) {G M_\bullet \over c^2} \gtrsim 10 {GM_\bullet \over c^2} . 
\end{equation}
In other words, an IMBH EMRI interpretation is at least allowed physically (for a single collision), as this tidal radius is outside of the direct capture radius of the black holes spacetime (and so the original TDE could have occurred). The temperature implied in the collisional  model for these parameters 
\begin{equation}
k T_{\rm BB} \approx 10 \, {\rm eV}, \quad \eta \approx 20, \quad \xi < 1, 
\end{equation}
would naively imply $kT_{\rm obs} \approx \eta^2 kT_{\rm BB} \approx 4 \, {\rm keV} \gg 100\, {\rm eV} = kT_{\rm data}$. However, at this level of photon starvation it is not clear if a simple $\eta^2$ scaling is sufficient to model the detailed radiative transfer, so it is unclear how meaningful this discrepancy is \citep[e.g., numerical radiative transfer simulations such as those performed by][would be of use here]{Vurm24}. 

As discussed above, we can however rule out this IMBH scenario by considering how much mass would be accreted onto the IMBH in each disk crossing. The Bondi accretion rate is given by 
\begin{equation}
\dot M_B \approx {4 \pi G^2 m_\bullet^2 \rho_{\rm disk} \over v_K^3}, 
\end{equation}
meaning that in one orbital crossing $t_{\rm cross} \approx h/v_K$, an amount of mass 
\begin{equation}
\Delta M_{\rm acc, \bullet} \approx f_B t_{\rm cross} \dot M_B \approx f_B {\pi \Sigma_{\rm disk} R_{\rm QPE}^2} \left({m_\bullet \over M_\bullet}\right)^2 , 
\end{equation}
where $f_B$ is an efficiency of mass accretion in units of the Bondi accretion rate. For the IMBH mass inferred above this is 
\begin{equation}
\Delta M_{\rm acc, \bullet} \approx 0.04 f_B M_{\rm disk}^{\rm QPE}, 
\end{equation}
meaning the entire disk would be drained within a timeframe of $\Delta t_{\rm acc} = P_{\rm QPE} M_{\rm disk}^{\rm QPE}/\Delta M_{\rm acc, \bullet} \simeq 50 / f_B$ days. As AT2019qiz has already been observed to show QPEs for 700 days \citep{Nicholl24} this final model is also ruled out. 

We stress again that this is a general property of IMBH models. To reach the observed energy scale of QPE flares (Figure \ref{fig:data}) they must contain such massive black holes $m_\bullet \sim 10^5 M_\odot$ that they would completely drain the accretion flow within a few crossings. 

\subsubsection{Collisions with extended debris streams}
The above analysis shows that collisions between spherical objects (either stellar surfaces or an IMBH) cannot reproduce observations of QPEs in TDEs. A Hills sphere model of collisions can be quickly ruled out as the ratio $E_{\rm rad, Hills}/E_{\rm obs}$ and $t_{\rm diff, Hills}/t_{\rm obs}$ both depend on the TDE stellar parameters to the same power, and so the ratio of these two quantities only depends on $(M_\star^{\rm emri}/M_\odot)^{2/9}$. The observed ratio of these two factors for AT2019qiz is $\sim 8.3$, a ratio that would require $M_\star^{\rm emri}/M_\odot \sim 10^4$, clearly implausible. 

We therefore move to models describing the collision between a debris stream (with width in the disk plane set either by $\sim R_\star^{\rm emri}$ or $\sim R_H$) which is vertically elongated. Note that the requirement for vertical elongation can be derived here on completely energetic grounds, and is in effect independent of the results of \cite{Yao24} (who showed that such a debris stream seems to form in numerical simulations of star-disk collisions). 

Irrespective of the cross section colliding in the disk plane, the length of the QPE flare is, in this limit, set by the vertical extension of the debris stream, and is broadly given by 
\begin{equation}
    {\Delta t} \approx 0.5\, {\rm hours}\,  \left({P_{\rm QPE} \over 4\, {\rm hours}}\right)^{4/3} \left({M_\star^{\rm emri} \over M_\odot}\right)^{1/2}\left({R_\star^{\rm emri} \over R_\odot}\right)^{-1/2}\left({M_\bullet \over 10^6M_\odot}\right)^{-1/3} ,
\end{equation}
for AT2019qiz this evaluates to 
\begin{equation}
    {\Delta t} \approx 11\, {\rm hours}\,  \left({M_\star^{\rm emri} \over M_\odot}\right)^{1/2}\left({R_\star^{\rm emri} \over R_\odot}\right)^{-1/2},
\end{equation}
% meaning that the mass to radius ratio of the EMRI must equal 
% \begin{equation}
%     {M_\star^{\rm emri} \over R_\star^{\rm emri}} \approx {7 M_\odot \over 10 R_\odot } .
% \end{equation}
% Again, assuming a simple mass-radius relationship of the form $R_\star^{\rm emri} = R_\odot (M_\star^{\rm emri} / M_\odot)^{4/5}$, we find 
% \begin{equation}
%     M_\star^{\rm emri} \approx 0.15 M_\odot, \quad R_\star^{\rm emri} \approx 0.2 R_\odot.
% \end{equation}
which means that for reasonable stellar parameters of the EMRI this flare timescale is perfectly plausible (especially given the variability in the flare timings of AT2019qiz \citealt{Nicholl24}, and the uncertainty in the above expression). %Feeding this into the radiated energy we find 
% \begin{equation}
%     {E_{\rm obs}\over E_{\rm rad}} = Q^{1/2} \approx {f_d^{1/2}\over \beta^{1/4}} \left({M_\star^{\rm tde}\over M_\odot}\right)^{5/12} \left({R_\star^{\rm tde} \over R_{\odot}}\right)^{1/4} , 
% \end{equation}
% where $Q^{1/2} \approx {600}$ for the Hills-debris stream model, and $Q^{1/2}\approx 300$ for the debris stream only model. Neither produces a workable solution for the properties of the star which was disrupted. 

To see if such an EMRI model could work, we take the simplifying assumptions that $f_d=1, \beta=1, h/r=0.1$ and that both stars follow the same main sequence mass-size relationship $R_\star/R_\odot = (M_\star / M_\odot)^{4/5}$. This leads to 
\begin{equation}
    {E_{\rm obs}\over E_{\rm rad}} = A = \left({M_\star^{\rm tde}\over M_\odot}\right)^{37/60} \left({M_\star^{\rm emri}\over M_\odot}\right)^{17/10} \approx 16, 
\end{equation}
for the Hills sphere-debris stream case, and 
\begin{equation}
    {E_{\rm obs}\over E_{\rm rad, debris}} = A' = \left({M_\star^{\rm tde}\over M_\odot}\right)^{37/60} \left({M_\star^{\rm emri}\over M_\odot}\right)^{19/30} \approx 78, 
\end{equation}
for the star-debris stream case. This second (star + debris stream) case is unworkable, but the first can be satisfied by plausible (if slightly massive) stellar parameters. Indeed, if both stars were $M_\star^{\rm tde} \approx M_\star^{\rm emri} \approx 3 M_\odot$, then the observed energetics of AT2019qiz could be explained. We remind the reader that various $\sim {\cal O}(1)$ parameters have been dropped in the analysis of this paper, and so these masses are not definitive, although their rough scales are.  We note that this large a stellar mass (for the disrupted star) is in contention with the mass inferred from the light curve fitting of the TDE itself \citep{Nicholl24}, but of course uncertainties exist in such modeling as well. 

The predicted blackbody temperature of the radiation in this limit would be 
\begin{equation}
    kT_{\rm BB} \approx 6.6 \,{\rm eV},
\end{equation}
and the photon starvation parameter 
\begin{equation}
    \eta \approx 0.2, 
\end{equation}
i.e., the emission would be too soft (according to this model of the post-shock evolution). It is not clear if this is tells us that this final model cannot reproduce the data, or whether the radiative transfer physics of the post-shocked debris is more complicated than that put forward here \citep[using the framework of][]{Linial2023}. This second option, that simple estimates of the temperature of the flares are insufficient to capture the evolution of the spectra, is supported by the simulations of \cite{Vurm24} who found $kT\sim 100$ eV emission in their simulation with not much sensitivity to other parameters.  Within the models put forward here however, the temperature of the emission is quite broadly speaking difficult to reconcile with the observations of all QPE sources, as we discuss in more detail below. 

% We can be somewhat more model independent in this analysis in fact, and not rely on any assumptions of the evolution of the post-shocked gas for AT2019qiz. Even in the limit at which all of the kinetic energy imparted into the gas from the collision between a TDE disk and an EMRI on an orbit corresponding to a QPE period of 48 hours 
% \begin{equation}
%     E_{\rm kinetic} \approx 1.3 \times 10^{45} \, {\rm erg} \,  {f_d\over \beta^{1/2}} \left({M_\star^{\rm tde}\over M_\odot}\right)^{5/6}   \left({R_\star^{\rm tde} \over R_{\odot}}\right)^{1/2}  \left({R_\star^{\rm emri} \over R_{\odot}}\right)^2 , 
% \end{equation}
% were turned into photons with 100\% efficiency, all of which were then observed, this would still be difficult to reconcile with the observed energy in each flare of AT2019qiz
% \begin{equation}
%     E_{\rm flare, \, AT2019qiz} \approx 6 \times 10^{47} {\rm erg} .
% \end{equation}
% In other words, even in the absence of any losses (observable or physical), a collision at this radius between an object and a disk formed following a TDE is a very low energy collision (relatively speaking), which will ultimately force all models which rely on such collisions to produce extreme parameters for the colliding body to reproduce the observations. 

\subsection{ The observed temperatures of QPE flares for $\big(P_{\rm QPE} \sim {\cal O}({\rm hours})\big)$ QPEs }
The main difference between the results of TDE disk collisions, and steady-state AGN-like disk collisions, is that TDE disks are significantly more dense at small radii when the disk first becomes intercepted by an object (as a $\sim$ stars worth of material was dumped on short radial scales). The main implication of this result is that it is more likely that the shock-heated gas resulting from the collision will be able to produce enough photons (a process that scales like $\sim \rho_{\rm disk}^{15/16}$ for Bremsstrahlung) to reach thermodynamic equilibrium, and therefore produce a blackbody emission spectrum. This can be seen by comparing the QPE-period dependence of the photon starvation parameter $\eta$ implied by a TDE disk model $\eta \propto P_{\rm QPE}^{25/24}$ with that of \cite{Linial2023}, $\eta \propto P_{\rm QPE}^{-49/24}$ (note that this is independent of the properties of the orbiting body). This means that for a TDE disk the emission should escape with temperature $kT_{\rm BB}$, which is typically much softer $kT_{\rm BB} \sim 10\, {\rm eV}$ than observed $kT_{\rm data} \sim 100\, {\rm eV}$. 

One way around this problem, while remaining within the framework put forward by \cite{Linial2023}, could be utilising the fact that at times significantly later than when the disk first spreads to the QPE-interception radius $R_{\rm QPE}$ this density profile will be modified, provided that the dominant mass change in the TDE disk system remains accretion (although this is unclear, as the disk evolution may be stalled by stripping the mass off the colliding star, see above). A mass accretion rate through the inner edge $\dot M_{\rm acc} \propto t^{-n}$, coupled with a spreading disk outer edge $R_{\rm out} \propto t^{2n-2}$, leads to a density drop with time of $\Sigma_{\rm disk} \propto t^{-n}$ (at fixed radius, see Appendix \ref{app:evolve} for a proof). This would lead to gradual heating of the blackbody temperature of the shocked gas $T_{\rm BB} \propto t^{n/4}$, and an increase in photon starvation $\eta \propto t^{9n/8}$, and decreasing cooling effects of inverse-Comptonization $y \propto t^{-n/16}$. It is possible therefore that short-period QPE sources involve a disk formed from an old TDE, and have evolved to the point where the observed temperature becomes hot enough to be detectable $kT_{\rm obs} \propto \eta^2 kT_{\rm BB} \sim t^{5n/2}$. 

This explanation would, however, lead to an interesting and eminently testable prediction, which may help determine whether photon starvation is indeed the solution to the temperature problem, or if new physics is required. As the peak temperature in the accretion flow itself (the one which produces the quiescent flux in between eruptions) decays with time as $T_{\rm disk} \sim t^{-n/4}$, with the same evolutionary timescale,  one should have 
\begin{equation}
    {T_{\rm QPE}(t_2)\over T_{\rm QPE}(t_1)} = \left({T_{\rm disk}(t_1) \over T_{\rm disk}(t_2)}\right)^{10} ,
\end{equation}
an extremely sensitive dependence which should be clearly detectable given a long enough observing baseline. We note that the best candidate for long time QPE disk evolution (i.e., the oldest source, GSN 069) evolves so slowly \citep{Miniutti2023, Guolo25}, that a $T_{\rm QPE}\sim (T_{\rm disk})^{-10}$ dependence would only change $T_{\rm QPE}$ by a factor $\sim 2$ \citep{Guolo25}. 

We believe that the temperature scaling of collisional models (or more broadly the spectral evolution of the flares) should be the subject of future study, with numerical simulations like those of \cite{Vurm24} rerun with the lower density disks likely to form following a TDE, and the extended debris structure discussed here. This is important because while the energetics of collisions between extended debris structures and a TDE disk can produce enough energy, it is not yet clear if the spectral evolution will be similar to observed QPE systems.

\subsection{ The duty-cycle of QPE sources }
A problematic element of most forms of the TDE disk-EMRI collision model is the predicted $t_{\rm QPE}-P_{\rm QPE}$ scaling relationship, namely 
\begin{align}
&t_{\rm QPE} \propto P_{\rm QPE}^{-2/3} \quad ({\rm ``Hard\, sphere" \, stellar\,\, EMRI}), \\ 
&t_{\rm QPE} \propto P_{\rm QPE}^0 \quad ({\rm Black\,\, hole\,\, EMRI}), 
\\
&t_{\rm QPE} \propto P_{\rm QPE}^{0} \quad ({\rm ``Hills\, sphere" \, stellar\,\, EMRI}), \\
&t_{\rm QPE} \propto P_{\rm QPE}^{+4/3} \quad ({\rm ``Stellar\, debris\, stream" \,  EMRI}),  
\end{align}
where we split off ``hard'' stellar EMRIs (i.e., the object has a collisional size given by its radius), and ``softer'' stellar EMRIs (i.e., the object has a collisional size set by either its Hills sphere or the size of the debris stream). 

All models except the extended debris stream are in contention with the data, which is well described by a constant duty cycle relationship  $t_{\rm QPE} = {\cal D} P_{\rm QPE}$, with ${\cal D} \approx 0.25$ \citep[][see Figure \ref{fig:data}]{Nicholl24}. The prediction of a negative (or at best flat for the IMBH and Hills mass case) scaling between $t_{\rm QPE}$ and $P_{\rm QPE}$ is an inevitable result of collisional models which do not have significant vertical  extension due to a simple fact about TDE disks: they have a fixed mass budget set by the mass of the disrupted star, and so objects intercepting this disk at larger radii (orbital periods) must intercept a disk with a smaller density, as this finite mass budget will have been spread over a larger area. As the time for the emission to escape from the shock-heated gas is set by the optical depth of the gas cloud (in all cases where there is not significant vertical extension), and therefore by the mass of the ejecta, this lower density inevitably means it is easier, and quicker, for photons to escape for larger QPE periods. 

Note that this argument is independent of, but broadly in agreement with (assuming the collisional paradigm is correct) the finding in \cite{Yao24}, who showed that vertically extended debris streams do form from collisions with accretion flows.  

\subsection{The energy of QPE flares}
\begin{figure}
    \centering
    \includegraphics[width=0.75\linewidth]{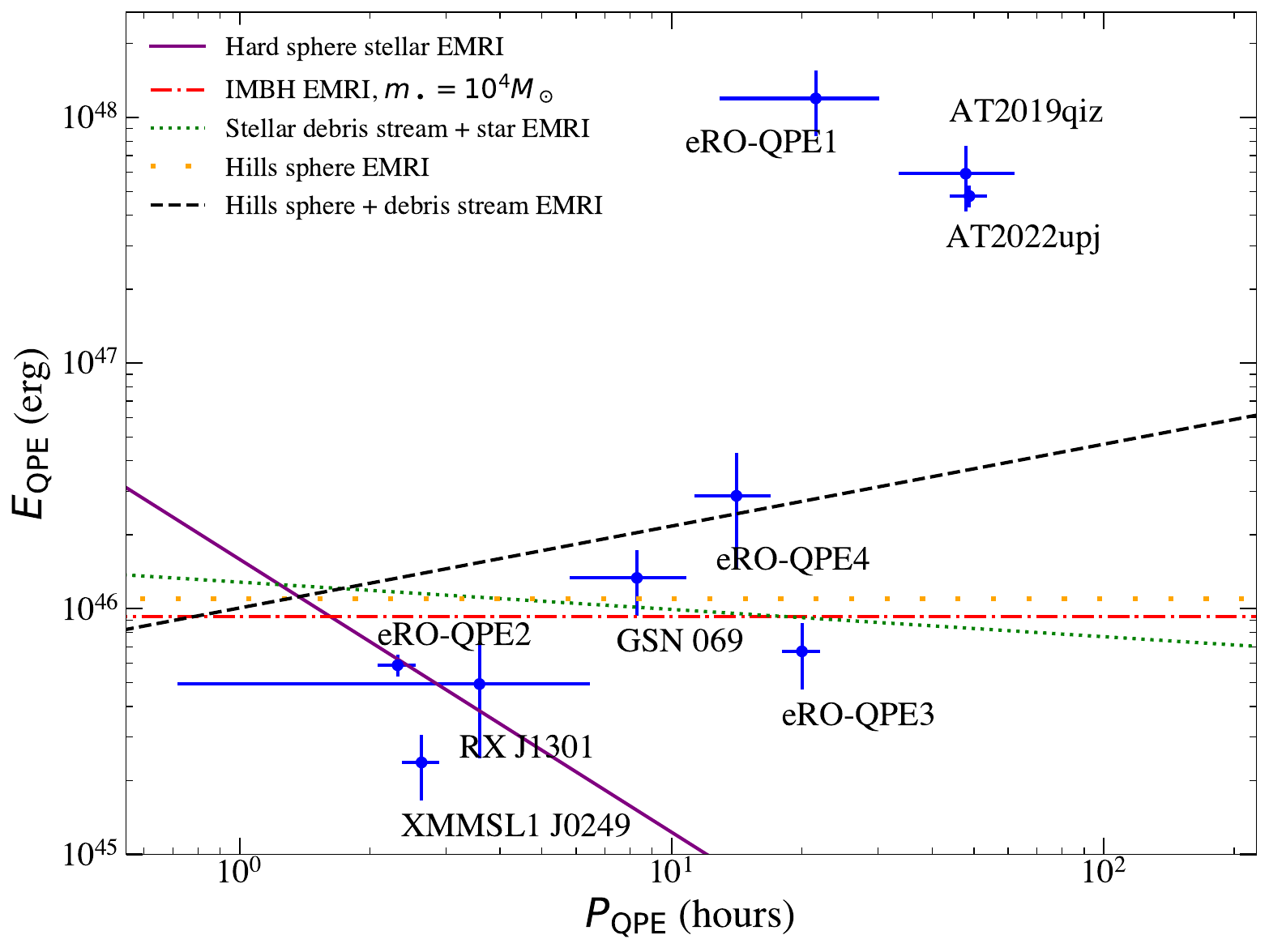}
    \caption{The different energy scales of collisional QPE models developed here, as a function of orbital period, with all other parameters set equal to their default values (i.e., all stellar properties are solar, $f_d=\beta=1$, $h/r=0.1$ and $M_\bullet = 10^6 M_\odot$). The hard sphere stellar EMRI model can be ruled out on energetic grounds for nearly all QPEs, while the IMBH model is pathological as it would result in the entire disk mass being accreted (by the secondary black hole) on $\sim$ year timescales. The various models of extended debris structures provide a reasonable description of low-energy QPE energetics, but only the Hills sphere + debris stream model can reach the high energy systems (and does require super solar $\sim 3M_\odot$ stellar properties). Generally speaking, it is difficult, but not impossible, for collisional models to reach the largest energy scales observed in QPE systems.  }
    \label{fig:models}
\end{figure}

As discussed throughout this work, the hardest element of QPE observations to explain within the collision model paradigm is the observed $E_{\rm rad}-P_{\rm QPE}$ scaling relationship, and in particular the highest energy QPE sources with $E_{\rm flare} \sim {\rm few} \times 10^{47}$ erg. The predicted scalings of the five models developed in this paper are  
\begin{align}
&E_{\rm rad} \propto P_{\rm QPE}^{-10/9} \quad ({\rm ``Hard\, sphere"\, stellar\,\, EMRI}), \\ 
&E_{\rm rad} \propto P_{\rm QPE}^0 \quad ({\rm Black\,\, hole\,\, EMRI}), \\
&E_{\rm rad} \propto P_{\rm QPE}^{-1/9} \quad ({\rm ``Stellar\, debris\, stream\, and \, star"\,  EMRI}), \\ 
&E_{\rm rad} \propto P_{\rm QPE}^{0} \quad ({\rm ``Hills\, sphere"\, stellar\,\, EMRI}), \\ 
&E_{\rm rad} \propto P_{\rm QPE}^{+1/3} \quad ({\rm ``Stellar\, debris\, stream\, and \, Hills\, sphere"\,  EMRI}), 
\end{align}
which are all somewhat in contention with the data (Figure \ref{fig:data}), but to varying degrees. Indeed, we showed in this paper that only one of these models can explain the highest energy flares of the QPE population (with non-pathological parameters), namely those flares seen in AT2019qiz. This model is the final one listed above, consisting of a puffed up star which is filling its Hills sphere, followed by a stream of its own stellar debris. We note that other models (just a debris stream or just a Hills sphere) are acceptable descriptions for the lowest energy QPEs (Figure \ref{fig:models}). We reiterate the point made earlier that IMBH models cannot reproduce any of the observations, as the IMBH mass required to sweep up enough matter in its Bondi radius to reproduce the energetics ($m_\bullet \sim 10^5 M_\odot$) would also accrete $\sim 1\%$ of the disk mass on each passing. 

This difficulty in explaining the observations can be traced back to the energy in the initial collision itself, which must drop with QPE period as less mass is left in larger TDE disks (a result of angular momentum conservation constraints), and the orbiting body is moving more slowly. This initial kinetic energy scales as
\begin{align}
    &E_{\rm ej} \propto P_{\rm QPE}^{-7/3}, \quad {\rm (``Hard\, sphere"\, stellar \, EMRI)}, \\
    &E_{\rm ej} \propto P_{\rm QPE}^{-1}, \quad {\rm (Black\, hole \, EMRI)}, \\ 
    &E_{\rm ej} \propto P_{\rm QPE}^{-4/3}, \quad {\rm (``Stellar\, debris\, stream\, and\, star" \, EMRI)}, \\
    &E_{\rm ej} \propto P_{\rm QPE}^{-1}, \quad {\rm (``Hills\, sphere"\, stellar \, EMRI)}, \\
    &E_{\rm ej} \propto P_{\rm QPE}^{-2/3}, \quad {\rm (``Stellar\, debris\, stream\, and\, Hills\, sphere" \, EMRI)} .
\end{align}

\section{Conclusions}\label{conc}
In this Paper we have extended the EMRI collision model of \cite{Linial2023} to include models of quiescent disks formed in the aftermath of a TDE and more complicated collisional geometries. The extension to TDE disks is required owing to the various links between TDEs and QPEs, culminating in the discovery of the sources AT2019qiz and AT2022upj, which are observed to have QPEs following otherwise ordinary TDEs.   This new model is in general over-constrained (if the black hole mass is known), with 3 observable properties set by two physical quantities: the total effective collisional area of the EMRI-disk collision (which may be the size of the object, or as we have shown is more likely to be a debris stream scale), and a product of quantities relating to the star which was tidally disrupted. 

These extended models can be expressed entirely in terms of the properties of the two stars involved in the process (the one disrupted and the one on an EMRI), and the mass of the black hole. These new models are therefore much more readily testable, as these parameters must satisfy known plausibility bounds (i.e., they must invoke high mass stars only rarely). 

For the particular source AT2019qiz both a simple stellar EMRI  and an IMBH can be immediately ruled out from the data as they cannot generate enough energy in their collisions (except for extremely high IMBH masses). Indeed, both models struggle to reproduce any observations of QPE systems (e.g., Figure \ref{fig:models} and previous discussion). 

On purely energetic grounds the only parameter space of collisional models not ruled out by AT2019qiz is that of collisions between extended debris structures and a TDE disk, with the debris structure needing to be extended both in the disk plane (out to $\sim$ its Hills radius) and also in the vertical direction (to an extended debris stream).  This vertical extension would then set the time of the flare, and the combined horizontal and vertical extension would (just) allow enough mass to be swept up for $\sim 3M_\odot$ stars (both the EMRI and TDE would need to be quite massive). We note that, of course, there are various $\sim {\cal O}(1)$ factors by which the calculations presented here could be (and likely are) wrong, meaning the precise masses are not set in stone. We do not however believe that these order unity factors will change the fact that the highest energy QPE flares can only be described (within the collision paradigm) by such extended structures, or the broader scaling relationships of the energetics calculated here.

This difficulty in reproducing the observed energies of QPE flares is a somewhat surprising result, as numerical simulations of the spectral evolution of the disk-orbiter collision model are in good accord with observations \citep{Vurm24}, although these simulations used a model of the quiescent disk (and colliding body) which we argue here differs markedly from the TDE case. The collisional model also successfully predicted that QPEs should switch on when the disk in a TDE spread (from the initial compact disk at $\sim 2r_T$) by angular momentum conservation to a large radius which coincides with the orbital radius, exactly as found in AT2019qiz \citep{Nicholl24}, AT2022upj \citep{Chakraborty25}, and eRO-QPE2 \citep{Wevers25}.  Although we do note that the disk in GSN 069 was large enough to have collision powered QPE flares in 2014, when no flares were detected \citep{Guolo25}.

We leave whether or not this last region of collision model parameter space can reproduce the observations of the full population of QPE sources as an open question, as to answer it will require numerical simulations (with radiation) of the collision process for these modified disk parameters and collision geometries. We believe that the models developed here will be an important first-order point of comparison, and in particular the structure of the disk (and how it differs markedly from steady state models) will be an important point for future studies. We do believe however that various order one factors will need to be verified by numerical models.  

The following four problems should be tackled by numerical simulations 
\begin{enumerate}
    \item Can repeated collisions between a star and a disk with TDE densities produce an extended debris structure (width $\sim$ Hills radius, and a larger height) with sufficient debris density to eject disk material upon the next collision? 
    \item Can collisions between this extended debris structure and low density disk material reach high enough radiated energies to explain observations for typical stellar parameters for the EMRI and disrupted star? 
    \item Can collisions between this extended debris structure and low density disk material reach high enough temperatures to explain observations? 
    \item What are the long time evolutionary states of these systems? This is especially relevant for short orbital periods where many evolutionary timescales appear comparable. 
\end{enumerate}
We believe that attempts to resolve this high-radiated energy tension should pay particular attention to the case of AT2019qiz, which has likely the best constrained disk properties of all QPE sources, while being the most difficult set of observations to reproduce with collisional models. 

%We note in passing that if one chooses by hand the scaling $R_{\rm collision} \sim R_{\rm QPE}^{5/2} \sim P_{\rm QPE}^{5/3}$ then one recovers $t_{\rm QPE} \sim P_{\rm QPE}$, $E_{\rm rad} \sim P_{\rm QPE}^{5/3}$ and $L_{\rm QPE} \sim P_{\rm QPE}^{5/9}$ all in reasonable accord with observations. As far as the author is aware however, there is no known mechanism to engineer such a scaling. 

\section*{Acknowledgments}
I am extremely grateful to Itai Linial, Brian Metzger and Elliot Quataert for discussions about this work, which significantly improved the paper. I am grateful to Muryel Guolo for comments on an earlier draft. 

This work was supported by a Leverhulme Trust International Professorship grant [number LIP-202-014]. For the purpose of Open Access, AM has applied a CC BY public copyright license to any Author Accepted Manuscript version arising from this submission. 
 
\section*{Data accessibility statement}
The data used in Figure \ref{fig:data} is publicly available.

\bibliographystyle{mnras}
\bibliography{andy}

\appendix

\section{Other surface density profiles}\label{app}
The results presented in the main body of this paper assume that the surface density of the disk at the location at which an orbiting body intercepts it is given by $\Sigma_{\rm out} \approx M_{\rm disk}/\pi R_{\rm out}^2$, i.e., it assumes a constant surface density in the disk at all radii. In this Appendix we show that changing this assumption introduces only small (order unity) changes to the amplitude of various quantities, and does not modify the scaling relationships derived in the main body of the paper. 

Assume instead that the disk surface density is described by 
\begin{equation}
    \Sigma(R) = \Sigma_{\rm out} \left({R\over R_{\rm out}}\right)^p, 
\end{equation}
then 
\begin{equation}
    M_{\rm disk} = 2\pi \int_{R_{\rm in}}^{R_{\rm out}} R \Sigma(R)\, {\rm d}R ,
\end{equation}
implies 
\begin{equation}
    \Sigma_{\rm out} =  {M_{\rm disk} \over \pi R_{\rm out}^2} \left[{2 + p \over 2(1 - x^{2+p})}\right], 
\end{equation}
where $x \equiv R_{\rm in}/R_{\rm out}$. As the properties of $M_{\rm disk}$ and $R_{\rm out}$ remain fixed by the orbit of the secondary and the properties of the star which was initially disrupted, the parameter scaling of this surface density at the point of interception is unchanged from the main body of the text. The multiplicative factor $(2+p)/(2(1-x^{2+p}))$ introduces only small changes to the amplitude of various quantities, depending slightly on the value of $p$. The value of $p$ depends on ones choice for the disk stress within an $\alpha$ framework, and is generally poorly constrained. 

Similarly, the total angular momentum of the disk 
\begin{equation}
    J_{\rm disk} = 2\pi \int_{R_{\rm in}}^{R_{\rm out}} R \Sigma(R)\, \sqrt{GM_\bullet R}\, {\rm d}R ,
\end{equation}
becomes 
\begin{equation}
    J_{\rm disk} =  M_{\rm disk} \sqrt{GM_\bullet R_{\rm out} } \left[{(4 + 2p)(1-x^{(5+2p)/2}) \over (5+2p)(1 - x^{2+p})}\right], 
\end{equation}
from which the disk angular momentum constraint used in the main body of the paper can be found in the limit $p=0, x\to 0$. 
\section{Constraints on the scale height of the disk}\label{app:height}
The disk scale height is given by the solution of the equations of vertical hydrostatic equilibrium, namely 
\begin{equation}
    {{\partial P} \over \partial z} \approx - {GM_\bullet \rho z \over r^3},
\end{equation}
the leading order solution of which is 
\begin{equation}
    H = \sqrt{P r^4 \over \rho J^2 }, 
\end{equation}
where $J$ is the disk specific angular momentum at a given radius (formally $J^2 = U_\phi^2 - a^2c^2(U_t^2 -1 )$ in relativity where $U_\phi$ and $U_t$ are the specific angular momentum and energy of the disk fluid, and $a$ is the black hole spin parameter \citep{Abramowicz97}). In the Newtonian limit $J^2 = GM_\bullet r$. Then, using $\rho = \Sigma/H$, we have
\begin{equation}
    H = {Pr^3 \over GM_\bullet \Sigma_{\rm disk} } = {r^3 \over GM_\bullet \Sigma_{\rm disk} } \left({k_B \Sigma_{\rm disk} T_c \over \mu m_p H} + {4\sigma T_c^4 \over 3 c}\right),
\end{equation}
where we have included both radiation and gas pressures, and introduced the disk central temperature $T_c$, the mass of a proton $m_p$, and the average mass of a particle in the disk $\mu$. This central temperature is related to the effective temperature of the disk by $T_c \approx (\kappa_{\rm es} \Sigma_{\rm disk})^{1/4} T_{\rm eff}$, where $T_{\rm eff}$ is the temperature of the radiation field that escapes the disk and $\kappa_{\rm es}$ is the electron scattering opacity.  This is a simple quadratic to be solved for $H$, namely 
\begin{equation}
H = {r^4 \over J^2 }\left({2\sigma T_c^4 \over 3 \Sigma_{\rm disk} c}\right) + \sqrt{{r^8 \over J^4} \left({2\sigma T_c^4 \over 3 \Sigma_{\rm disk} c}\right)^2 + \left({k_B T_c \over \mu m_p}\right){r^4 \over J^2} }  ,
\end{equation}
or equivalently 
\begin{equation}
H = {r^4 \over J^2 }\left({2\sigma \kappa_{\rm es} T_{\rm eff}^4 \over 3 c}\right) + \sqrt{{r^8 \over J^4} \left({2\sigma \kappa_{\rm es} T_{\rm eff}^4 \over 3  c}\right)^2 + \left({k_B T_{\rm eff} (\kappa_{\rm es} \Sigma_{\rm disk})^{1/4} \over \mu m_p}\right){r^4 \over J^2} }  .
\end{equation}
If one can measure the temperature of the inner disk from X-ray observations, which we shall denote $T_{\rm eff, max}$, then one can use the constraints of energy conservation to relate it to the effective temperature of the disk at the collision radius $R_{\rm QPE}$, namely
\begin{equation}
    T_{\rm eff} = T_{\rm eff, max} \left({R_{\rm QPE} \over R_I}\right)^{-3/4} ,
\end{equation}
where $R_I \sim 6 GM_\bullet/c^2$ is the innermost stable circular orbit radius (the typical radial scale of the hottest disk material). This means that with an observation of $T_{\rm eff, max}$ and an estimate of $M_\bullet$, one can compute $H$ at the collisional radius. 

Finally, we note that this expression neglects magnetic pressure support (i.e., there should be a term $P_{\rm mag} = B^2/8\pi$ in the above equation for hydrostatic equilibrium). If the magnetic pressure of the disk exceeds the scale of both the thermal and radiation pressures in the disk \citep[as is sometimes assumed in disk analyses, e.g.,][]{Begelman07} then this will naturally lead to order unity changes to the above expression.  

\section{Evolution of the flare properties if accretion dominated}\label{app:evolve}
Let us assume that the evolution of the system is dominated by continual accretion (i.e., $\dot M_{\rm eject}, \dot M_{\rm abl} \ll \dot M_{\rm acc}$), then the evolution of the flare properties would be dominated by the dropping surface density of the flow. 

The accretion disk continues to lose mass through it's inner edge via 
\begin{equation}
    \dot M_{\rm acc} \approx - (n - 1) {M_{\rm disk, 0} \over t_{\rm visc}} \left({t \over t_{\rm visc}}\right)^{-n}  \propto t^{-n} ,
\end{equation}
where $n$ is {\it apriori} unknown, as it depends on the details of the turbulent transport of angular momentum through the disk, and the normalisation is chosen so that all of the mass is accreted in the limit $t\to \infty$. This expression is valid only at times $t \gtrsim t_{\rm visc}$. We know that $n > 1$ so that all of the mass is eventually accreted. The mass of the disk therefore drops like 
\begin{equation}
    M_{\rm disk} = M_{{\rm disk}, 0} - \int_{t_{\rm visc}}^t \dot M_{\rm acc}(t')\, {\rm d}t' \propto t^{1-n} ,
\end{equation}
where the initial disk mass cancels owing to the normalisation of the accretion rate. This causes (as we have discussed throughout this paper) the outer edge of the disk to grow with time to conserve angular momentum 
\begin{equation}
    J_{\rm disk} \propto M_{\rm disk} \sqrt{R_{\rm out}} \approx {\rm const} \to R_{\rm out} \propto t^{2n-2} .
\end{equation}
Therefore 
\begin{equation}
    \Sigma_{\rm disk}(R=R_{\rm QPE}) \propto {M_{\rm disk} \over R_{\rm out}^2}\left({R_{\rm QPE} \over R_{\rm out}}\right)^p \propto t^{-(5+2p)(n-1)} .
\end{equation}
We have added in a radial dependence $\Sigma \propto R^p$ here, because it now becomes important what radial dependence the disk density has (for example all of the mass could relatively quickly move past the EMRI orbiting radius and the collisions be occurring between the less dense inner flow). Note that in conventional $\alpha$ theory, the index $p$ is entirely specified by the index $n$, and in fact we have\footnote{Technically this assumes a disk scale height which depends only on radius.} \citep{LBP74,Pringle91}
\begin{equation}
    p = {5-4n\over 2n-2}  \quad \to \quad (5+2p)(n-1) = n .
\end{equation}
 We then simply ``tag'' the various powers of surface density in all of the various observable quantities 
\begin{align}
    M_{\rm ej} &\propto \Sigma \propto t^{-n} , \\
    E_{\rm ej} &\propto \Sigma \propto t^{-n} , \\
    t_{\rm diff} &\propto \Sigma^{1/2} \propto t^{-n/2} , \\
    E_{\rm rad} &\propto \Sigma^{1/2} \propto t^{-n/2} , \\
    T_{\rm BB} &\propto \Sigma^{-1/4} \propto t^{n/4} , \\
    \eta &\propto \Sigma^{-9/8} \propto t^{9n/8} , \\
    y_{\rm max} &\propto \Sigma^{1/16} \propto t^{-n/16} . 
\end{align}
These time dependence's hold irrespective of the nature of the orbiting body (i.e., ``stellar debris'', ``hard sphere'' or black hole EMRI). Note that $t_{\rm diff}$ only sets the QPE flare length in models which do not include an extended vertical debris stream. As the vertical debris stream extent sets the time dependence of the flare in some of the models (star + debris stream and Hills sphere + debris stream). These models will have 
\begin{equation}
    L_{\rm flare} \propto  \Sigma^{1/2} \propto t^{-n/2} ,
\end{equation}
while the hard-sphere, IMBH, and Hills sphere models will have a constant flare luminosity
\begin{equation}
    L_{\rm flare} \propto \Sigma^0 \propto t^{0} . 
\end{equation}

\end{document}